\documentclass{jfm}
\usepackage{natbib}
\usepackage{graphicx}
\graphicspath{{figures/}} 
\usepackage{amsmath} 
\usepackage{amsfonts}
\usepackage{amssymb}
\usepackage{bm}
\usepackage{mathbbol}
\usepackage{xparse}
\usepackage{ifthen}	
\usepackage{xspace}
\usepackage{hyperref}	
\usepackage{siunitx}
\usepackage{multirow}
\usepackage{comment}
\usepackage{array}
\usepackage{gensymb}
\usepackage{subfigure}
\usepackage[usenames,dvipsnames]{xcolor}
\usepackage[section]{placeins}
\usepackage{soul}

\shorttitle{Direct cascade in forced two-dimensional turbulence}
\shortauthor{M. Reynoso, D. Zhigunov, and R. O. Grigoriev}

\title{Self-similarity and the direct (enstrophy) cascade in forced two-dimensional fluid turbulence}

\author{Mateo Reynoso, Dmitriy Zhigunov,
\and Roman O. Grigoriev
\corresp{\email{roman.grigoriev@physics,gatech.edu}}}

\affiliation{School of Physics, Georgia Institute of Technology, 
Atlanta, GA 30332, USA}

\begin{document}

\maketitle

\begin{abstract}
In the absence of large-scale coherent structures, a widely used statistical theory of 2D turbulence developed by Kraichnan, Leith, and Batchelor (KLB) predicts a power-law scaling for the energy, $E(k)\propto k^\alpha$ with an integral exponent $\alpha={-3}$, in the inertial range associated with the direct cascade. A power-law scaling is also observed in the presence of coherent structures, but the scaling exponent becomes fractal and often differs substantially from the value predicted by the KLB theory. Here we present a dynamical theory that sheds new light on the relationship between the spatial and temporal structure of the large-scale flow and the scaling of small-scale structures representing filamentary vorticity. Specifically, we find hyperbolic regions of the large-scale flow to play a key role in the flux of enstrophy between scales. Small-scale vorticity in these regions can be described by dynamically self-similar solutions of the Euler equation, which explains the power-law scaling. Furthermore, we find that correlations between different hyperbolic regions are responsible for the emergence of fractal scaling exponents.
\end{abstract}


\section{Introduction}

Fluid turbulence is a paradigm unsolved problem of classical physics and mathematics. It occupies a unique place in science due to its complicated chaotic and multiscale nature and in engineering due to its ubiquity and practical importance. The origins of turbulent cascades, which are responsible for generating structure at multiple scales are among the key mysteries. In particular, in two-dimensional (2D) turbulence, there are two different cascades: direct and inverse \citep{clercx2009,boffetta2012}. Specifically, the inverse cascade transfers energy from smaller to larger scales, while the direct cascade transfers enstrophy from larger to smaller scales. Here we focus on fluid flows driven at large scales, for which it is the direct cascade that is primarily responsible for the multiscale nature of turbulence.

The first theoretical, statistical description of both the inverse and the direct cascade was developed by \citet{kraichnan1967}, \citet{leith1968} and \citet{batchelor1969}. In particular, for the direct cascade, the Kraichan, Leith and Batchelor's (KLB) theory predicts the energy density in the Fourier space to scale as a power law,  $E(k)\propto k^{-3}$. Experiments \citep{rivera2014} and numerical simulations \citep{herring1985,legras1988,maltrud1991} however generally find the spectrum to be steeper, with $E(k)$ exhibiting scaling close to a power law $k^\alpha$ with a non-integral exponent in the range $-5 < \alpha < -3 $. The deviations from the predictions of the KLB theory, which assumes the small scales to be uncorrelated, have been attributed to the presence of coherent structures, which introduce correlations. Indeed, the integral scaling exponent $\alpha=-3$ can be recovered if the coherent structures, and hence the correlations, are artificially destroyed \citep{benzi1988,maltrud1991,borue1993,chen2003}. The importance of coherent structures, which reflect the accumulation of energy at large scales caused by the inverse cascade, has been recognized already by \citet{kraichnan1971} whose logarithmic correction to the $k^{-3}$ power law reflects the nonlocal nature of the direct cascade. 

While no systematic description of the effects of coherent structures has been developed so far, several different mechanisms that may contribute to the direct cascade have been explored. Earlier studies have focused on the stretching of small patches of vorticity, essentially treated as a passive scalar, by large-scale vortices, leading to filamentation. In particular, \citet{saffman1971} predicted the exponent $\alpha=-4$ by using a simplified picture which assumed that the vorticity is uniform inside each patch and the corresponding vorticity field has a finite number of discontinuities along any straight line. Corresponding scaling, however, is only observed at early times, while on longer time scales both assumptions break down and the spectrum often becomes less steep \citep{brachet1987}. It was hypothesized that, on longer time scales, vorticity filaments are stretched as they are wound up around vortices, generating a fractal structure that is characterized by a nonintegral exponent. For point vortices, the corresponding exponent was predicted to be $\alpha=-11/3=-3.6(6)$ \citep{moffatt1986,gilbert1988}. While these results established a relationship between the topology of the large-scale flow and the scaling exponent, they did not explain the lack of universality observed in the presence of coherent structures. Indeed, coherent structures feature vortices with a wide variety of shapes and hence stretching properties \citep{zhigunov2022}.

More recent studies have identified straining (or hyperbolic) regions of large-scale flows, rather than vortices (or elliptic regions), as playing a key role in the direct cascade. Numerical results of \citet{chen2003} confirmed the conjecture of \citet{kraichnan1971} that the primary physical mechanism behind the direct cascade is ``vortex thinning,'' i.e., steepening of the vorticity gradients in strain-dominated regions of the flow. This picture is supported by the experimental studies of \citet{kelley2011} and \citet{liao2015} which found that, at small scales, the spectral fluxes of both the energy and the enstrophy are enhanced in regions that are predominantly straining. This relation is particularly clear in numerical simulations of body-forced turbulence on a square doubly periodic domain where the large-scale flow at high Reynolds numbers ($Re$) is dominated by a pair of counter-rotating vortices, and vorticity filaments are mostly found in the straining regions \citep{zhigunov2022}. The presence of pronounced coherent structures in this flow leads to particularly strong deviations from the KLB predictions, with the scaling exponent reaching values as high as $\alpha\approx-5$. While convincing, this evidence is qualitative, and a quantitative theory is yet to be developed that can predict the scaling exponents found in both numerics and experiments and explain their connection to the spatial and temporal properties of coherent structures.

This paper introduces the key elements of such a quantitative theory for bounded flows that reach statistical equilibrium in the presence of driving and viscous dissipation. The paper is structured as follows. We will start by analyzing the large- and small-scale structures of vorticity that emerge in direct numerical simulations of high-$Re$ 2D flows in Section \ref{sec:dns}. Section \ref{sec:results} introduces a dynamical model for small-scale vorticity as well as several classes of self-similar solutions representative of observations and discusses their properties. The results are discussed in Section \ref{sec:discussion} and conclusions are presented in Section \ref{sec:conclusions}.

\section{High-$Re$ flow in two dimensions}\label{sec:dns}

To gain intuition into the physical mechanism of the direct cascade, it is helpful to analyze typical vorticity structures that emerge in high-$Re$ incompressible flows driven by steady large-scale forcing on a doubly periodic domain ($0\leq x,y<2\pi$). Such flows can be described by the Navier-Stokes equation
\begin{align}\label{eq:NSE}
    \partial_t{\omega}+{\bf u}\cdot\nabla{\bf \omega}=\nu\nabla^2{\omega}+\varphi,
\end{align}
written in terms of vorticity $\omega = \partial_x u_y - \partial_y u_x = - \nabla^2 \psi$, where $u_x = \partial_y \psi$ and $u_y = -\partial_x \psi$ are the velocity components, $\psi$ is the stream function, and $\varphi$ describes the external forcing. In this section, we will focus on the results of numerical simulations of turbulent flow with $\nu=10^{-5}$ driven by a steady checkerboard forcing $\varphi = \Phi\sin(k_fx)\sin(k_fy)$ with $k_f=4$ reported by \citet{zhigunov2022}. The forcing wavenumber was chosen to be sufficiently low to make sure there is a large separation between the forcing length scale $\ell_f=k_f^{-1}$ and the Taylor microscale $\ell_t=Re^{-1/2}\ell_0$ at which viscous effects become important. Here, the Reynolds number is defined as $Re=\mathcal{E}\ell_0^2/\nu$, where $\mathcal{E}$ is the characteristic magnitude of velocity gradient tensor elements $\partial_i u_j$ and $\ell_0$ is the characteristic length scale of the large-scale flow, so that $\ell_t= (\nu/\mathcal{E})^{1/2}$. To make sure the small-scale structures are resolved, the flow was computed on a grid with resolution $2048\times 2048$. Accounting for the 2/3 dealiasing used in the numerical simulations, this corresponds to the highest resolved wavenumber of $k_{max}=682$.

\begin{figure}
    \centering
       \subfigure[]{\includegraphics[height=0.4\textwidth]{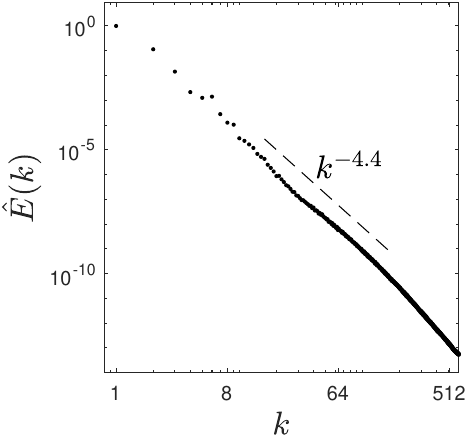}}
       \hspace{1mm}
       \subfigure[]{\includegraphics[height=0.4\textwidth]{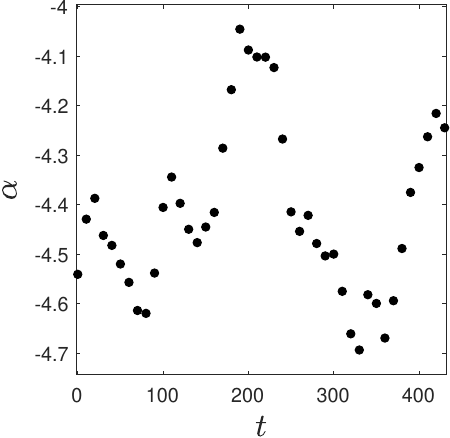}}
    \caption{The energy of a turbulent flow. (a) The energy spectrum averaged over a long time interval ($10^3$ nondimensional units). (b) The exponent of the spectrum is shown at intervals of the characteristic time scale $T_c=10$, averaged with nearby points.}
    \label{fig:spectrum}
\end{figure}

The large-scale flow is found to be insensitive to the particular choice of the forcing profile, so long as its frequency is relatively low \citep{kim2010,kim2015,kim2017}. This is due to the accumulation of the energy at the largest scales accessible to the flow, as illustrated by \autoref{fig:spectrum}(a), caused by the inverse cascade. The energy spectrum averaged over a long time interval is found to exhibit a power-law scaling $E(k) \propto k^\alpha$ over roughly a decade in the wavenumber ($16 \leq k \leq 170$). This scaling indicates the presence of an inertial range characteristic of fully developed turbulence. The exponent $\alpha\approx -4.4$ is found to be quite different from that predicted by KLB theory. Numerical simulations also suggest that $\mathcal{E}\propto \Phi/(k_f\nu)$, which implies the balance between the forcing and viscous dissipation terms and, consequently, the dominant balance between the terms $\partial_t{\omega}$ and ${\bf u}\cdot\nabla{\bf \omega}$ at high Reynolds numbers.

\begin{figure}
    \centering
    \subfigure[]{\includegraphics[height=0.4\textwidth]{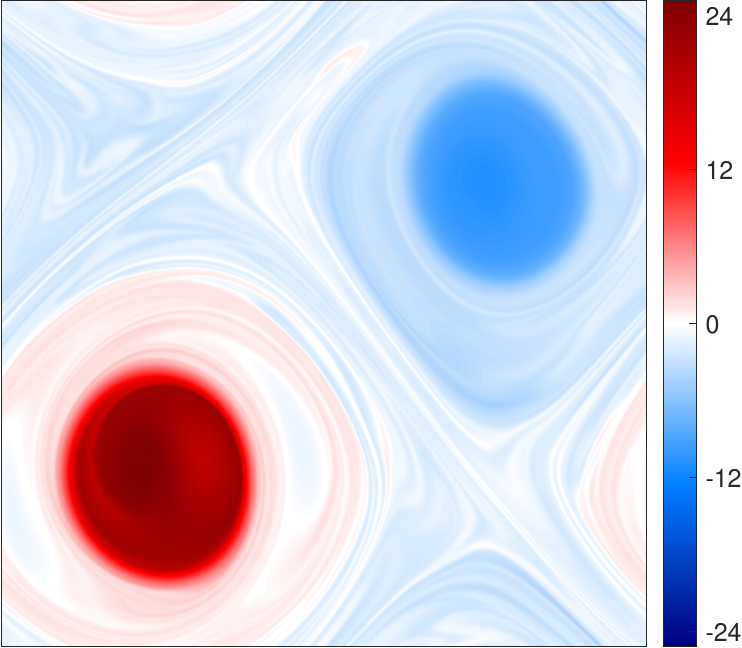}}\hspace{5mm}
    \subfigure[]{\includegraphics[height=0.4\textwidth]{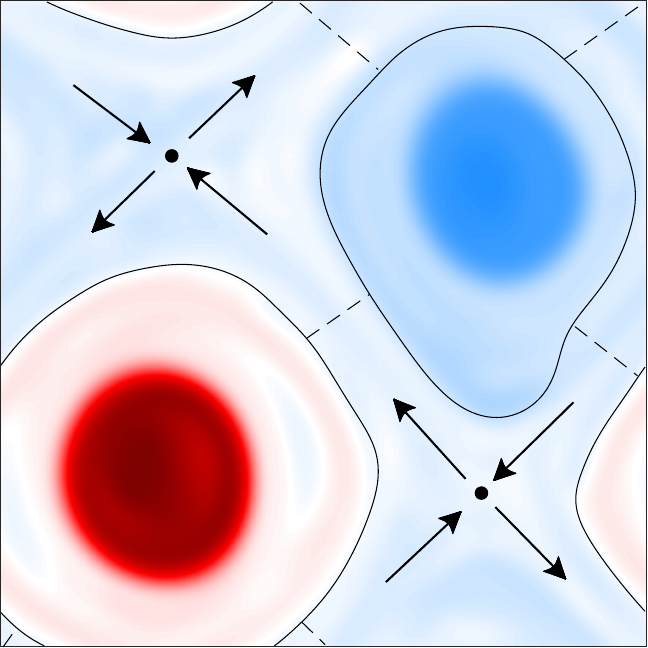}} \\
    \subfigure[]{\includegraphics[height=0.4\textwidth]{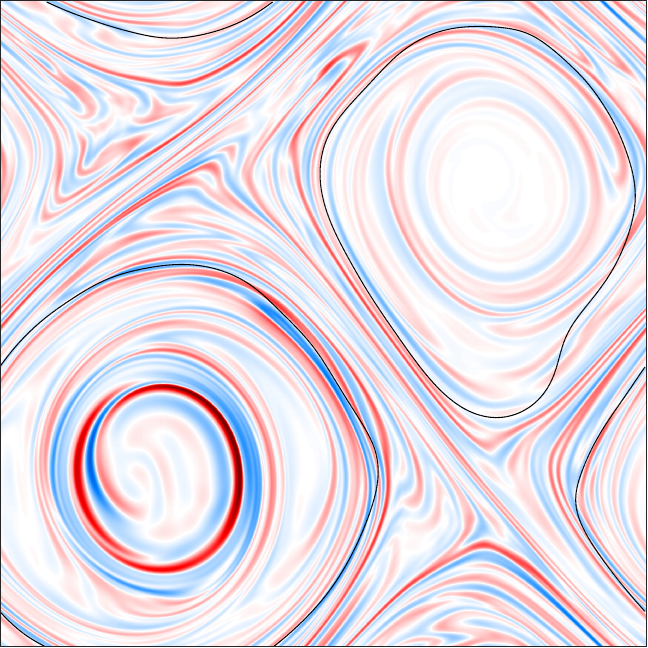}}\hspace{10mm}
    \subfigure[]{\includegraphics[height=0.4\textwidth]{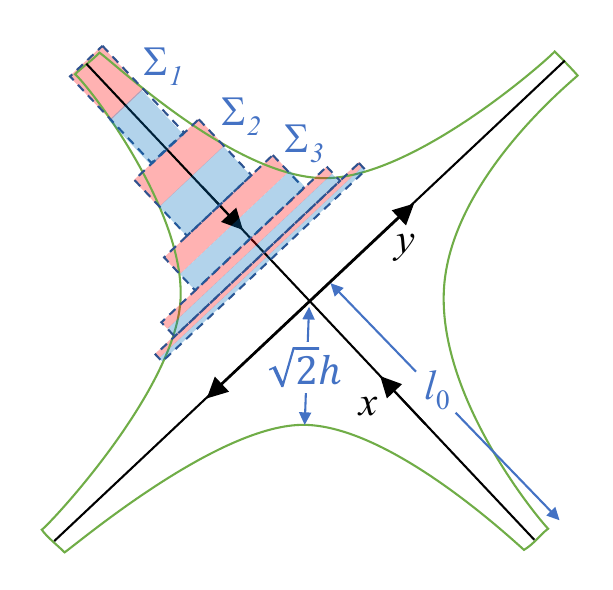}}
    \caption{Typical snapshot of a High-$Re$ turbulent flow. (a) Vorticity field $\omega$. (b) The large-scale component of $\omega$ (corresponding to wavenumbers $k\le 16$) features two pronounced counter-rotating vortices. The boundaries between regions dominated by vortical and straining flow are shown as solid black lines and the boundaries between the two hyperbolic regions are shown as dashed black lines. The arrows indicate the direction of the flow in each hyperbolic region and the corresponding stagnation points are shown as black circles. The boundaries of the hyperbolic regions, stagnation points, and direction of the flow are shown in a reference frame co-moving with the large-scale vortex pattern. (c) The small-scale component of $\omega$ (corresponding to wavenumbers $k\ge 16$) features pronounced vorticity filaments. (d) The local coordinate system and geometry of a hyperbolic region of the flow. The color map in all four panels uses red (blue) shades to indicate positive (negative) values of vorticity.}
    \label{fig:flow}
\end{figure}

Figure \ref{fig:flow}(a) shows a typical snapshot of the turbulent flow. We decomposed this flow into a small-scale and large-scale component by applying a filter that corresponds to a smoothed circular window with radius $2\sqrt{2}k_f$ in Fourier space, so the entire inertial range is included in the small-scale component. 
Several features of this flow are worth pointing out. The large-scale flow shown in Figure \ref{fig:flow}(b) is organized by a pair of counter-rotating vortices.
The region outside the vortex cores is dominated by the straining flow, with the strain driven by the two vortices. The region of straining flow has nearly uniform large-scale vorticity. The large-scale flow is not static, rather, a large fraction of the time, it is found to be nearly time-periodic \citep{zhigunov2022}. Although time-dependent, the flow structure remains largely unchanged, with both vortices executing a nearly circular motion. There are four stagnation points, a pair of elliptic ones inside the vortex cores and another pair of hyperbolic ones inside the region of straining flow outside of the vortex cores. 
When averaged over the characteristic time scale $T_c\approx 10$, which corresponds to the approximate period of the large-scale flow, the energy spectrum retains a power-law shape. However, the associated exponent $\alpha$ is found to vary over a significant range of values, as  \autoref{fig:spectrum}(b) shows, reflecting the changes in the properties of the large-scale flow on time scales longer than $T_c$. 

These results appear to be quite representative. Similar large-scale coherent structures in the form of ``vortex crystals'' \citep{aref2003} are also found to form for flows forced at small scales \citep{smith1994,chertkov2007}, even though they tend to take a rather long time to emerge. Notably, this behavior characteristic of forced turbulence is quite different from that found for freely decaying turbulence \citep{mcwilliams1984,bracco2000} where large-scale coherent structures fail to form.

For the flow considered here, most of the small-scale vorticity has a filamentary structure with the filaments mostly confined to the region of straining flow, as illustrated by Figure \ref{fig:flow}(c). This region can be decomposed into a pair of hyperbolic regions, each one containing a hyperbolic stagnation point. Inside each hyperbolic region, vorticity filaments are mostly oriented along the expanding direction of the straining flow, although some filaments are found to be oriented along the contracting direction. 
Some filamentary vorticity is also found inside the vortex cores. These filaments are found predominantly in the regions where the gradient of the large-scale vorticity is the largest and are oriented transversely to this gradient. Hence, both the straining and vortical regions contribute to the direct cascade, with the former corresponding to the picture described by \citet{kraichnan1971} and the latter corresponding to the picture described by \citet{moffatt1986} and \citet{gilbert1988}. We will focus here on the straining region which contains most of the filamentary vorticity. 

\section{Results}\label{sec:results}

To understand how the large-scale coherent flow affects the direct cascade, below we construct a dynamical model for small-scale vorticity inside the hyperbolic regions and analyze several types of analytical solutions representative of vorticity filaments observed in the numerics. All of these solutions feature structure at different scales, however, their associated energy and enstrophy spectra are found to be non-universal and depend on the properties of the large-scale flow, both temporal and spatial.

\subsection{The dynamics of small-scale vorticity}\label{sec:selfsim}

In the inertial range $k_f\ll k\ll k_t$, where $k_t=1/(\ell_t s)$, the effects of both forcing and viscous dissipation can be ignored, so the flow is effectively described by the Euler equation
\begin{align}\label{eq:vort}
    \partial_t\omega+{\bf u}\cdot\nabla\omega=0.
\end{align}
As discussed in the previous section, inside each hyperbolic region, the large-scale flow has essentially constant vorticity. Consider a reference frame in which a hyperbolic stagnation point remains at the origin and an orthogonal coordinate system with the $y$ coordinate oriented along the expanding direction of the flow, as shown in \autoref{fig:flow}(d). 
The velocity ${\bf u}_l$ of the large-scale flow inside the hyperbolic region can be written as
\begin{align}
    {\bf u}_l = -\mathcal{E} x \hat{x} + \mathcal{E}(y + 2\cot(\theta) x) \hat{y},
\end{align}
where $\mathcal{E}$ is the strain rate and $\theta$ is the angle between the expanding and contracting direction, hence the large-scale vorticity in the hyperbolic region is simply $\omega_l=2 \mathcal{E}\cot\theta$. For the simulations considered here, the time-dependence of the large-scale flow manifests mostly as a slow overall drift of the pattern, hence the strain rate $\mathcal{E}>0$, the orientation of the $y$ axis, and the angle $\theta$ can be considered constant (or at least slowly varying). This limit corresponds to the adiabatic approximation used by \citet{lapeyre1999,lapeyre2001}.

Separating the flow into the large- and small-scale contributions, e.g., ${\bf u}={\bf u}_l+{\bf u}_s$, and substituting the result into the Euler equation yields
\begin{align}\label{eq:vorts}
    \partial_t\omega_s+{\bf u}_l\cdot\nabla\omega_s=-{\bf u}_s\cdot\nabla\omega_s.
\end{align}
This evolution equation shows that small-scale vorticity $\omega_s$ can be considered a passive scalar that is advected by the large-scale flow ${\bf u}_l$, as is commonly assumed \citep{kraichnan1971}, only when the term ${\bf u}_s\cdot\nabla\omega_s$ vanishes. Indeed, when the level sets of the streamfunction $\psi_s$ describing the small-scale component of the flow are aligned with the level sets of the small-scale vorticity $\omega_s$, the right-hand-side of Equation \eqref{eq:vorts} disappears, so small-scale vorticity effectively becomes a passive scalar. This will indeed be the case for vorticity filaments whose radius of curvature is large compared with their thickness. As Figure \ref{fig:flow}(c) illustrates, this condition is satisfied almost everywhere except for the small regions where the filaments are folded sharply. 

In practice, the expanding and contracting directions of the straining flow tend to be nearly orthogonal so, without loss of generality, we will assume $\theta = \pi/2$ in most of the discussion below (the general case is considered in the Appendix \ref{app:B3}). The corresponding streamfunction then becomes $\psi_l=-\mathcal{E}xy$ and the large-scale vorticity vanishes, $\omega_l=0$. 
Let $\ell_0$ define one half of the distance between adjacent hyperbolic stagnation points of the large-scale flow. Inside the hyperbolic region sketched in Figure \ref{fig:flow}(d),
\begin{align}\label{eq:xy}
    |xy|\lesssim h^2,\quad |x|,|y|<\ell_0,
\end{align}
bounded by the level sets of $\psi_l$, Equation \eqref{eq:vorts} reduces to
\begin{align}\label{eq:vort1}
    \partial_t\omega-\mathcal{E}x\partial_x\omega+\mathcal{E} y \partial_y \omega=0.
\end{align}
%
%
The shape (described by the characteristic thickness $h$) of the hyperbolic region depends on the properties of the large-scale flow. 
In practice, $h$ can be found by computing the shape of the Lagrangian coherent structure \citep{haller2015} which defines the boundary between the elliptic and hyperbolic region of the large-scale flow.
As illustrated by Figure \ref{fig:flow}(b), the hyperbolic region does not have to be reflection-symmetric with respect to either the expanding or contracting direction. We assume it to be symmetric, without loss of generality, to simplify the notations.
The general solution to the linear equation \eqref{eq:vort1} can be found using the method of characteristics
\begin{align}\label{eq:self-sim}
    \omega=\omega_s=g(\chi,\eta),
\end{align}
where $g$ are arbitrary functions of the similarity variables $\chi\equiv e^{\mathcal{E}t}x/\ell_0$ and $\eta\equiv e^{-\mathcal{E}t}y/\ell_0$. This family of solutions is dynamically self-similar, i.e., $\omega(e^{\mathcal{E}\tau}x,e^{-\mathcal{E}\tau}y,t)=\omega(x,y,t+\tau)$ for any $\tau$, although only the values of $\tau$ corresponding to multiples of the temporal period $T=2\pi/\Omega$ of the large-scale flow are of interest here. This is illustrated in Figure \ref{fig:flow}(d); the vorticity field inside the region $\Sigma_1$ is compressed along the $x$ direction and stretched in the $y$ direction by the same factor $\Lambda=e^{\mathcal{E}T}>1$ after one period and mapped onto the region $\Sigma_2$. Similarly, the region $\Sigma_2$ is mapped onto the region $\Sigma_3$, etc.

\subsection{Non-interacting hyperbolic regions}\label{sec:uniform}

The simplest such solutions depend on just one of the two similarity variables.
Consider, for instance, small-scale vorticity filaments oriented along the expanding direction, in which case
\begin{align}\label{eq:self-sim+}
    \omega=g_c^\pm(\chi),
\end{align}
where the superscript denotes the sign of $x$. This family of solutions describes vortex thinning associated with the stretching of filaments as they are advected towards the $y$ axis. The functions $g_c^\pm(\chi)$ are determined by the variation in the vorticity field at the ``entrances'' to the hyperbolic region where $x=\pm\ell_0$. 
Since vorticity at $x=\ell_0$ and $x=-\ell_0$ will generally be different, so $g_c^+(\chi)$ and $g_c^-(\chi)$ will also be different. For oscillatory $g_c^\pm(\chi)$, these self-similar solutions describe vorticity filaments whose characteristic thickness decreases exponentially fast, $\Delta x(t)\propto e^{-\mathcal{E}t}$, as they are advected towards the streamline $x=0$, representing a continuous flux of enstrophy from large to small scales.

The family of solutions \eqref{eq:self-sim+} describes non-interacting hyperbolic regions, i.e., vorticity fields entering one hyperbolic region that have not just left another hyperbolic region. In this case, the temporal and spatial profile of the vorticity field at the entrance to a hyperbolic region is defined by the forcing field $\varphi$, which acquires time-dependence due to the time-periodic overall motion of the hyperbolic regions. For forcing with low spatial frequency, the vorticity field at either entrance ($x=-\ell_0$ or $x=\ell_0$) to the hyperbolic region can be considered to vary slowly in the $y$ direction, i.e., $\omega(\pm\ell_0,y,t)\approx g_c^\pm(e^{\mathcal{E}t})$. For  the large-scale flow with temporal period $T$, the functions $g^\pm_c(\chi)$ have to be periodic, with period $\mathcal{E}T$. As a result, the vorticity field described by the solution family \eqref{eq:self-sim+} becomes both time-periodic and self-similar in the conventional sense, i.e., $\omega(\Lambda x,t)=\omega(x,t)$.

Self-similarity of the vorticity field implies that its Fourier spectrum $\hat{\omega}({\bf k},t)$ has a power-law scaling. Indeed, since $g_c^\pm(\chi)$ is periodic, it can be written in the form of a Fourier series, i.e., inside each hyperbolic region we have
\begin{align}\label{eq:invis_soln}
    \omega=\sum_{n=1}^{\infty}\bar{a}_n^\pm\cos[n s \ln |\chi|+\bar{\phi}_n^\pm],
\end{align}
where $\bar{a}_n^\pm$ are the Fourier amplitudes, $\bar{\phi}_n^\pm$ are the corresponding phases, and  $s=\Omega/\mathcal{E}=2\pi/\ln(\Lambda)$ is a key nondimensional parameter describing the temporal frequency of the large-scale flow. It is straightforward to show that $|\hat{\omega}({\bf k},t)|\propto k_x^{-1} \delta(k_y)$, and therefore the enstrophy spectrum in the inertial range should have a power-law scaling 
\begin{align}\label{eq:saffman}
    H(k)\propto \langle|\hat{\omega}({\bf k},t)|^2\rangle_t \propto k^{-2},
\end{align}
which corresponds to the scaling exponent $\alpha=-4$ for the energy $E(k)=k^{-2}H(k)$, consistent with the prediction of \citet{saffman1971}  (see Appendix \ref{app:B2}).
This result however does not account for two essential features of the vorticity field. First of all, viscosity becomes important in the dissipation range leading to attenuation of the thinnest vorticity filaments. Second, vorticity filaments are confined to the hyperbolic region, which introduces a (weak) dependence on the $y$ coordinate. We will consider the consequences of both effects next.  

In the presence of viscosity, self-similarity breaks down on length scales comparable to the Taylor microscale $\ell_t$ and vorticity field takes the following form 
\begin{align}\label{eq:visc_soln}
    \omega = \sum \limits_{n=1}^{\infty} a_n^\pm\cos\left[n s \ln |\chi| +\phi_n^\pm\right],
\end{align}
where, to leading order in $\ell_t/x$,
\begin{align}
    \phi_n^\pm(x)=\bar{\phi}^\pm_n-\frac{ns}{2}\frac{\ell_t^2}{x^2}
\end{align}
and
\begin{align}\label{eq:amp}
    a_n^\pm(x)=\bar{a}_n^\pm\exp\left[-\frac{n^2s^2}{2}\frac{\ell_t^2}{x^2}\right].
\end{align}
as shown in Appendix \ref{app:A}. Note that for $|x|\gg s\ell_t$, both the phases $\phi_n^\pm(x)$ and the amplitudes $a_n^\pm(x)$ become constant and the self-similar solution \eqref{eq:invis_soln} is restored. For $|x|\lesssim s\ell_t$, vorticity is exponentially strongly suppressed. The former (latter) range in the physical space corresponds to the inertial (dissipation) range in the Fourier space.
 
The effect of confinement to the interior of the hyperbolic region can be represented by imposing an envelope on the vorticity field which approaches zero outside of the hyperbolic region and unity inside. This envelope can be written as a function of the variable $\xi \equiv\chi\eta=xy/\ell_0^2\propto \psi_l$ (which is an invariant of the large-scale flow ${\bf u}_l$, i.e., $\partial_t\xi+{\bf u}_l\cdot\nabla\xi=0$) such as $e^{-\ell_0^4\xi^2/h^4}$, which yields
\begin{align}\label{eq:amp_hyp}
    a_n^\pm(x,y)=\bar{a}_n^\pm\exp\left[-\frac{n^2s^2}{2}\frac{\ell_t^2}{x^2}-\frac{\ell_0^4\xi^2}{h^4}\right].
\end{align}
Computing the enstrophy for this choice on the infinite plane (i.e., $\ell_0\to\infty$), one finds
\begin{align}\label{eq:klb}
     H(k)\propto k^{-1}
\end{align}
in the inertial range, as shown in Appendix \ref{app:B1}. Other appropriate choices of the envelope would yield the same power law with the value of the scaling exponent $\alpha=-3$ predicted by the KLB theory.

\begin{figure}
    \centering
    \subfigure[]{\includegraphics[width=0.32\textwidth]{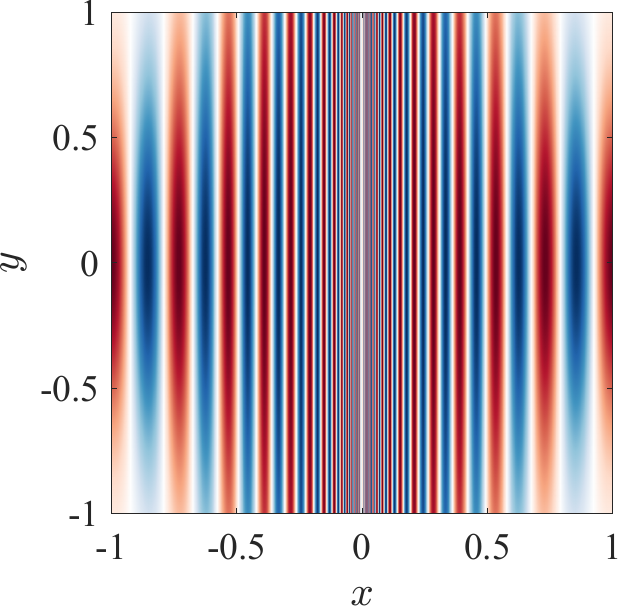}}\hspace{1mm}
    \subfigure[]{\includegraphics[width=0.32\textwidth]{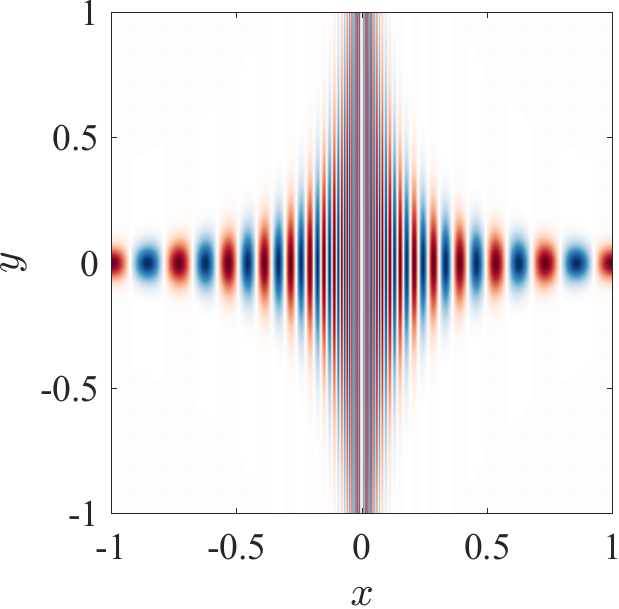}}\hspace{1mm}
    \subfigure[]{\includegraphics[width=0.32\textwidth]{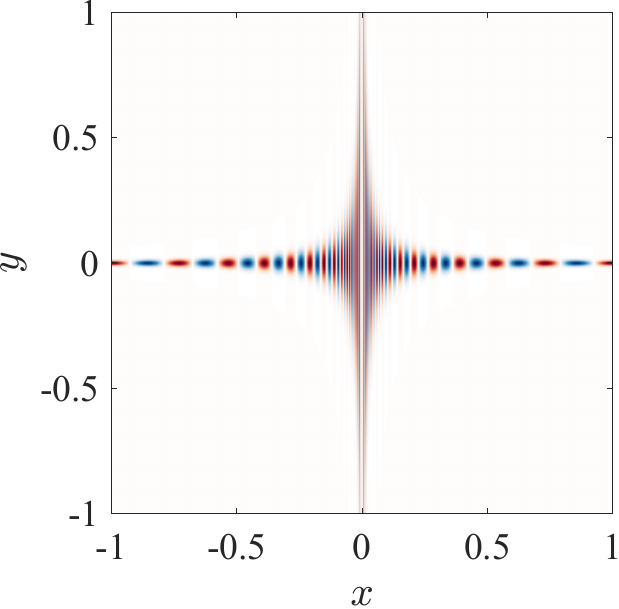}}\\
    \subfigure[]{\includegraphics[width=0.45\textwidth]{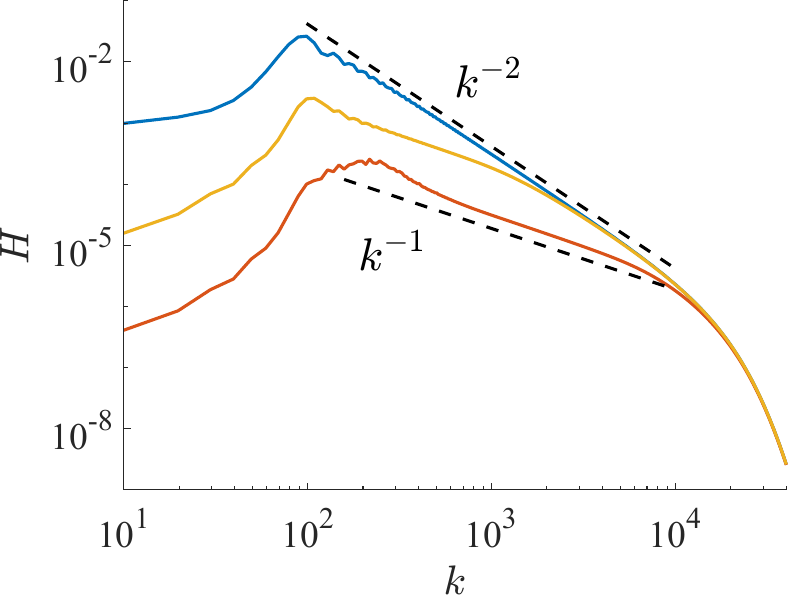}}
    \hspace{1mm}
    \subfigure[]{\includegraphics[width=0.45\textwidth]{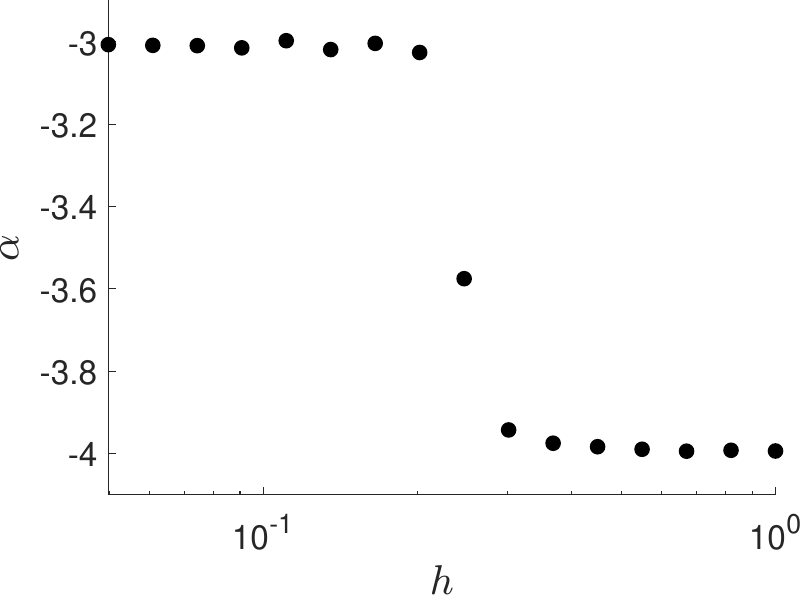}}
    \caption{Snapshots of the vorticity field \eqref{eq:omega1} for $s=20$, $\ell_0=1$, and $Re=10^7$ with (a)  
    $h=0.82$, (b) $h=0.25$, and (c) $h=0.1$. (d) The enstrophy spectrum shows different scaling regimes corresponding to $h=0.82$ (solid blue line) and $h = 0.25$ (solid yellow line). (e) Exponent $\alpha$ to describe the which corresponds to the best of a single-power law $H(k)\propto k^{\alpha+2}$ over the interval $k_c<k<k_t$ as a function of the hyperbolic region thickness $h$, where $k_c$ is defined by Equation \eqref{eq:kc}. 
    }
    \label{fig:2Dspec_h}
\end{figure}

Depending on the values of $h/\ell_0$ and $Re$, one could see one or both scalings \eqref{eq:saffman} and \eqref{eq:klb} in the inertial range, as illustrated in Figure \ref{fig:2Dspec_h}(d) for the special case $\bar{a}_n^\pm=\delta_{n1}$, i.e.
\begin{align}\label{eq:omega1}
    \omega=\cos(s\ln|\chi|)\exp\left[-\frac{s^2\ell_t^2}{2x^2}-\frac{\ell_0^4\xi^2}{h^4}\right],
\end{align}
with $\ell_0=1$ and $Re=10^7$, which corresponds to $k_t\approx 3\times 10^3$. Indeed, for thick hyperbolic regions (i.e., $h\gtrsim\ell_0$, cf. Figure \ref{fig:2Dspec_h}(a)), the envelope can be ignored, and we recover the scaling relation \eqref{eq:saffman}. For thin hyperbolic regions  (i.e., $h\ll\ell_0$, cf. Figure \ref{fig:2Dspec_h}(c)), the length of vorticity filaments is determined by the hyperbolic envelope rather than the size $\ell_0$ for $k$ in the inertial range, so we find the scaling relation \eqref{eq:klb} instead. For intermediate values of $h/\ell_0$ (cf. Figure \ref{fig:2Dspec_h}(b)), we can see both scaling regimes, $H(k)\propto k^{-1}$ at lower $k$ and $H(k)\propto k^{-2}$ at higher $k$. This is what one should expect: the hyperbolic envelope only affects the length of the thicker filaments while thinner filaments all have the same length $\ell_0$.

Observation of two different scaling regimes clearly requires the inertial range to be quite wide and, hence, $Re$ to be quite large. Asymptotic states at high $Re$ are likely inaccessible in experiments and require very long simulations to reach, which may explain the lack of experimental or numerical evidence showing two different scaling exponents.
For smaller $Re$, power-law scaling is only found over a narrow range of wavenumbers, and only one of the two regimes can be observed. Figure \ref{fig:2Dspec_h}(e) shows the exponent $\alpha$ for the best-fit power law $H(k)\propto k^{\alpha+2}$ as a function of the thickness $h$ of the hyperbolic region. We see that the exponent asymptotes to $-3$ for small $h$ and $-4$ for large $h$ with a narrow transition region where the fit to a single power-law becomes a poor approximation. 

The relation between the thickness of the hyperbolic region and the wavenumber at which the crossover between the two scaling regimes occurs can be easily estimated by inspecting \autoref{fig:flow}(d). The boundaries of the hyperbolic region $|xy|=h^2$ and $|y|=\ell_0$ intersect at positions $x=\pm x^*$, where $x^*=h^2/\ell_0$. The thickness of a vorticity filament entering the hyperbolic region at $x_0=\pm \ell_0$ is $\delta x_0\sim(1-\Lambda^{-1/2})\ell_0$. After $n$ periods, its position and thickness become $x_n=\pm\Lambda^{-n}\ell_0$ and $\delta x_n\sim\Lambda^{-n}(1-\Lambda^{-1/2})\ell_0=(1-\Lambda^{-1/2})|x_n|$, so the thickness of the filaments at $x=\pm x^*$ is $\delta x^*\sim(1-\Lambda^{-1/2})x^*$. The scaling exponent $\alpha=-3$ corresponds to $|x|>x^*$ (and hence $k<k^*$) while the scaling exponent $\alpha=-4$ corresponds to $|x|<x^*$ (and hence $k>k^*$), where
\begin{align}
    k^*=\frac{\pi}{\delta x^*}\sim\frac{\pi\ell_0}{(1-\Lambda^{-1/2})h^2}.
\end{align}
In particular, for $\ell_0=1$, $h=0.25$ and $s=20$ we find $\Lambda^{-1}=e^{-2\pi/s}\approx 0.73$ and $k^*\sim 345$, which is comparable to the value $k^*\sim 10^3$ found in \autoref{fig:2Dspec_h}(d).  

In comparison, for a typical DNS snapshot such as that shown in \autoref{fig:flow}(a), we find $\mathcal{E}\approx 3.7$, $\Omega\approx 0.63$, $\ell_0\approx 2.2$, and $h\approx 1$, yielding $k^*\sim 7$. The arguments presented in this section would suggest that the energy spectrum should scale as $E(k)\propto k^{-4}$ for $k>k^*$ and, indeed, we find a power law scaling in \autoref{fig:spectrum}(a) in this range of wavenumbers. However, the scaling exponent is found to be fractal, varying in time, and bounded by $\alpha=-4$, as \autoref{fig:spectrum}(b) illustrates. Of course, it would be a mistake to draw any far-reaching conclusion based on the analytical solution \eqref{eq:omega1}, which is missing several essential features of filamentary vorticity representative of turbulent flows, as discussed below.

Before we discuss the interaction effects, in conclusion of this section, for completeness, we should point out the existence of another family of self-similar solutions
\begin{align}\label{eq:self-sim-}
    \omega=g_e^\pm(\eta),
\end{align}
where $g_e^\pm(\eta)$ are again arbitrary functions and the superscript denotes the sign of $y$. This family of solutions describes vorticity filaments oriented along the contracting direction of the straining flow. These filaments become thicker, rather than thinner, with their characteristic thickness growing as $\Delta y(t)\propto e^{\mathcal{E}t}$, hence this family of solutions describes so-called backscatter or flux of enstrophy from small scales towards large scales. For a time-periodic large-scale flow, the corresponding small-scale vorticity field is also self-similar, $\omega(\Lambda^{-1} y,t)=\omega(y,t+T)$. Such solutions will only be dynamically relevant when the vorticity at the entrance of the hyperbolic region varies slowly in the $x$ and quickly in the $y$ direction, i.e., has the form of filaments oriented along the contracting direction. Indeed, vorticity filaments with such an orientation are routinely found in turbulent flows featuring pronounced coherent structures, as illustrated in Figure \ref{fig:flow}(c), and this is due entirely to the interaction between adjacent hyperbolic regions.

\subsection{Interacting hyperbolic regions}

\begin{figure}
    \centering
    \includegraphics[height=38mm]{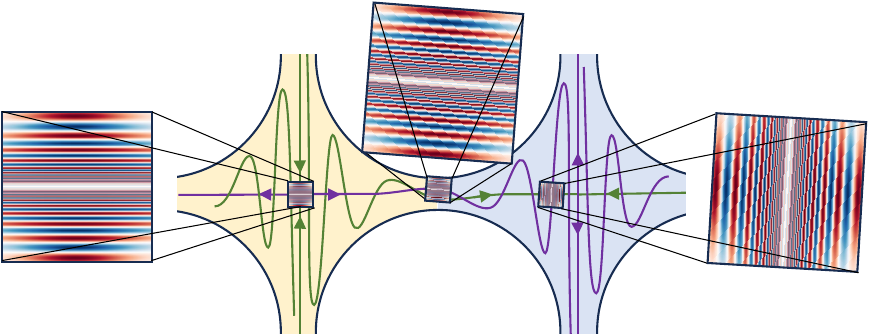}
    \caption{Vorticity transport between a pair of adjacent hyperbolic regions of a time-periodic large-scale flow. The unstable manifold (purple) of the saddle at the center of the left hyperbolic region (pale yellow) tangles with the stable manifold (green) of the saddle at the center of the right hyperbolic region (pale blue). Vorticity filaments everywhere are stretched along, and aligned with, the unstable manifold. }
    \label{fig:tangle}
\end{figure}

The large-scale flow associated with a pronounced coherent structure introduces strong correlations between adjacent hyperbolic regions: small-scale vorticity leaving one hyperbolic region immediately enters an adjacent one. The large-scale flow considered here (cf. Figure \ref{fig:flow}) is particularly simple since it contains just two hyperbolic regions with a saddle (fixed point or compact limit cycle) at the center of each. 
Each one of the two saddles of the large-scale flow has an associated stable and unstable manifold whose directions coincide, near the saddle, with the contracting and expanding direction of the large-scale flow. When the large-scale flow is time-periodic, the unstable manifold of one saddle crosses the stable manifold of the other, forming a heteroclinic tangle, as illustrated in Figure \ref{fig:tangle}, and leading to chaotic mixing \citep{ottino1990}. In fact, it is this chaotic mixing that is most likely responsible for the large-scale vorticity becoming effectively uniform inside the hyperbolic regions. The stretching of small-scale vorticity along the unstable manifold and compression transverse to it aligns vorticity filaments along the unstable manifold. Hence, it is such heteroclinic (or for some flows, homoclinic) tangles that determine the shape and size of the hyperbolic regions. The correlation between adjacent hyperbolic regions not only determines the orientation of the vorticity filaments at the entrance but also their Fourier spectrum.  This is illustrated in \autoref{fig:tangle} which shows what the vorticity field looks like qualitatively at different positions along the same unstable manifold. We will explore how both features, i.e., the orientation and the frequency spectrum of the filaments entering a hyperbolic region, affect the enstrophy spectra (in the $Re\to\infty$ limit) in the remainder of this section.

\subsubsection{The effect of the spatial orientation}

\begin{figure}
    \centering
    \subfigure[]{\includegraphics[height=0.3\textwidth]{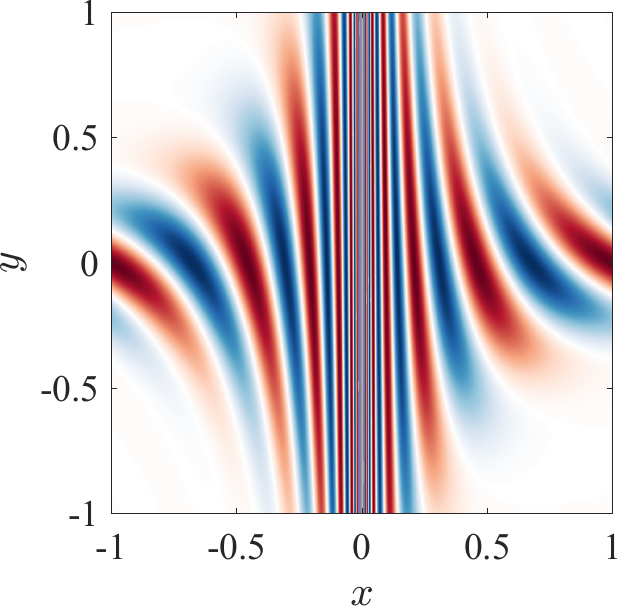}}\hspace{1mm}
    \subfigure[]{\includegraphics[height=0.3\textwidth]{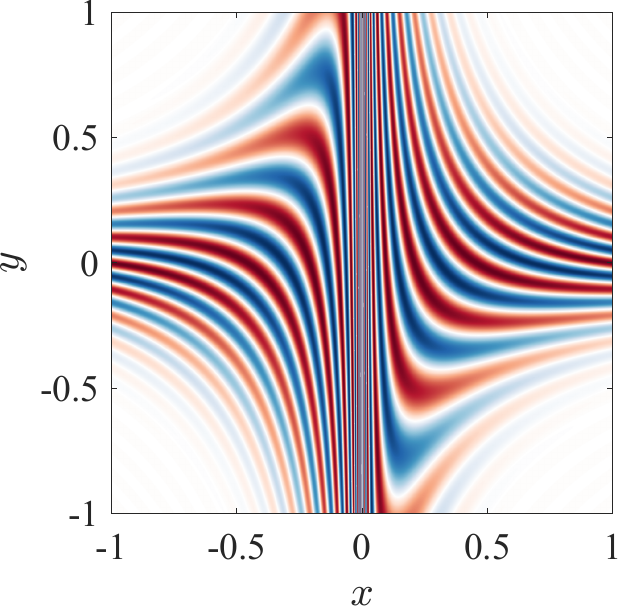}}\hspace{1mm}
    \subfigure[]{\includegraphics[height=0.3\textwidth]{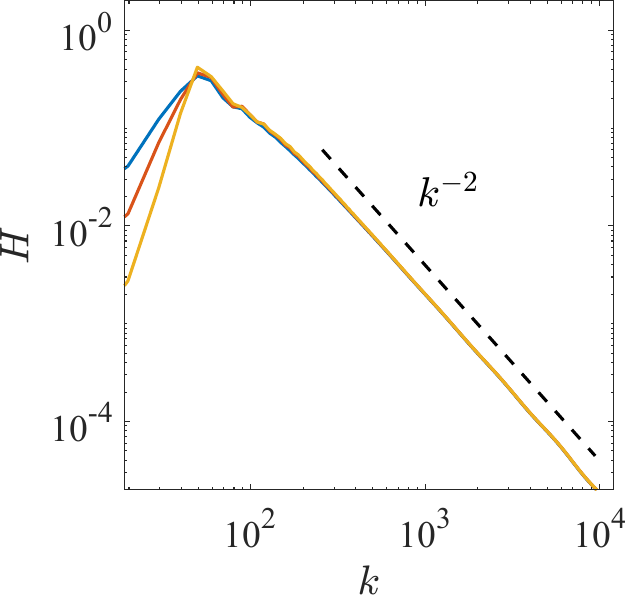}}
    \caption{Filaments of varying orientation at the entrance to the hyperbolic region. A snapshot of the vorticity field \eqref{eq:tilt_omega} for $s=8$, $h=0.5$, $\ell_0=1$ 
    and (a) $r = 1.5$ or (b) $r = 7.5$. (c) The enstrophy spectrum corresponds to the vorticity field for different orientations: $r = 2.5$ (blue), $r = 5$ (yellow), and $r = 7.5$ (red). 
    }
    \label{fig:tilted_soln}
\end{figure}

The solutions \eqref{eq:self-sim+} and \eqref{eq:self-sim-} may still be dynamically relevant even for the case of interacting hyperbolic regions when the vorticity filaments at the entrance are oriented strictly along the contracting or expanding direction of the large-scale flow. However, this is rather unlikely: their orientation is determined by the shape of the unstable manifold which generally will not be perfectly aligned with either direction. To determine whether the enstrophy and energy spectra retain a power-law shape and what the scaling exponent is for filaments with arbitrary orientations, let us consider a vorticity field
\begin{align}\label{eq:tilt_omega}
    \omega=\cos(s[\ln|\chi| + r\xi])
    e^{-\ell_0^4\xi^2/h^4},
\end{align}
which is a special case of the general solution \eqref{eq:self-sim}. Here $r$ is a non-dimensional parameter that controls the orientation of filaments at the entrance of the hyperbolic region. In particular, for $|r|\ll 1$ this solution reduces to \eqref{eq:omega1} in the inviscid limit, with the filaments at the entrance $x=\pm\ell_0$ oriented along the expanding direction ($y$ axis). For $|r|\gg 1$, the filaments become oriented everywhere along the streamlines of the large-scale flow, as illustrated in \autoref{fig:tilted_soln}(b). In particular, at the entrance, they become oriented along the contracting direction ($x$ axis). Note that the values of $r$ could be different on the opposite sides of the $y$ axis which is a separatix of the flow. Here we consider $r$ to have the same value for simplicity. Furthermore, note that the thickness of the filaments at the entrance is large and fairly uniform, with very thin filaments only found near the $y$ axis (i.e., close to the unstable manifold).

The enstrophy spectrum for this solution was computed numerically for different values of $r$, with representative results shown in \autoref{fig:tilted_soln}(c). We find that, regardless of the orientation (and curvature) of the filaments, the spectrum has the form of a power law with same (integral) exponent $\alpha=-4$ predicted for straight filaments oriented along the expanding direction of the large-scale flow in the limit of $h$-large. The same qualitative result (not shown) is found in the limit of $h$-small, where  $\alpha=-3$.
Hence, it is not the orientation of the vorticity filaments that is responsible for generating fractal exponents or the emergence of fractal structures in the vorticity field.

\subsubsection{The effect of the frequency content}\label{sec:fractal}

The small-scale vorticity field at the entrance to a hyperbolic region (e.g., the one shown in pale blue in \autoref{fig:tangle}) is determined by that at the exit from an immediately adjacent hyperbolic region (e.g., the one shown in pale yellow in \autoref{fig:tangle}). Recall that vorticity filaments are not just aligned with, but are also compressed in the direction transverse to, an unstable manifold. Hence, for a time-periodic large-scale flow, at the entrance to any hyperbolic region, one would generally find vorticity filaments with an entire spectrum of spatial frequencies, from extremely thin filaments near the unstable manifold of the adjacent saddle to much thicker filaments generated by the forcing $\varphi$ further away from the manifold. 

\begin{figure}
    \centering
    \subfigure[]{\includegraphics[height=0.3\textwidth]{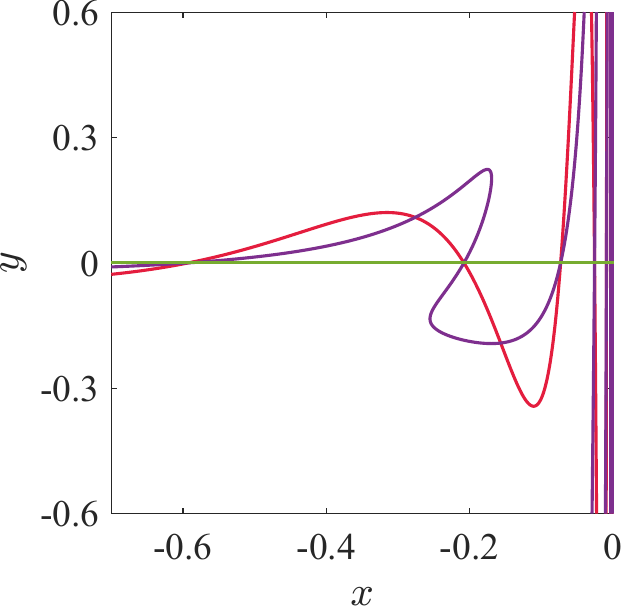}}\hspace{1mm}
    \subfigure[]{\includegraphics[height=0.3\textwidth]{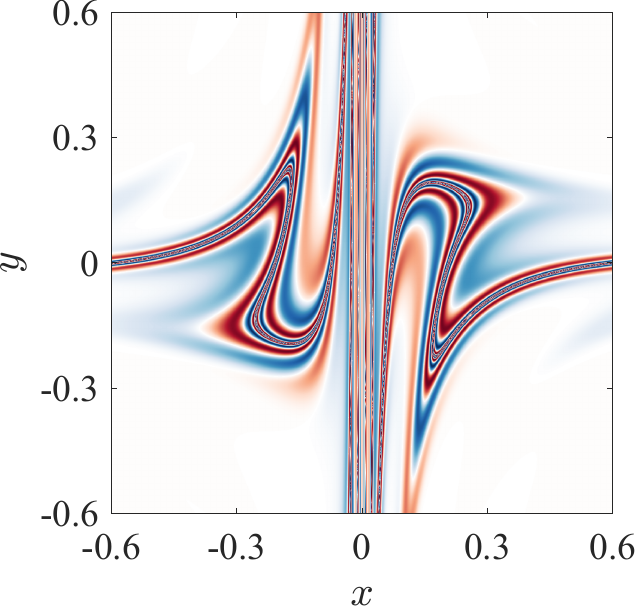}}\hspace{1mm}
    \subfigure[]{\includegraphics[height=0.3\textwidth]{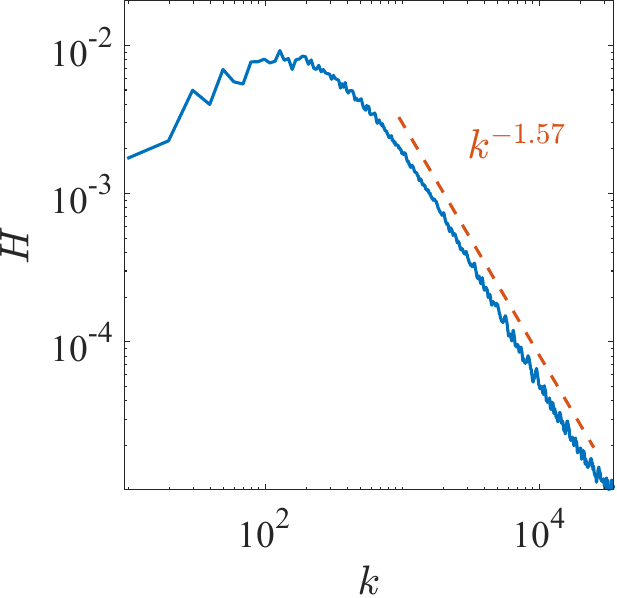}}
    \caption{Filaments with a broad spectrum of spatial frequencies at the entrance to the hyperbolic region. (a) The stable manifold (green) and the unstable manifold for $r = 0$ (red) and $r = 16$ (purple). (b) A snapshot of the vorticity field \eqref{eq:manf_soln} for $r = 16$.
    Enstrophy spectrum calculated by direct numerical evaluation as the time-average of \eqref{eq:manf_soln}, shown as a solid blue line (c). The best fit for a power-law is shown as a dashed red line. In all three panels, $s = 3$ and $h = 0. 2$.}
    \label{fig:fractal_soln}
\end{figure}

To construct a particular solution \eqref{eq:self-sim} for the vorticity field with such properties consider a function
\begin{align}\label{eq:manf}
    &q = \cos(s[\ln|\chi| - r \xi ]) - \frac{\ell_0^2\xi}{h^2}
\end{align}
whose level set $q=0$ defines an unstable manifold of an adjacent saddle. It intersects the stable manifold $y=0$ of the saddle $(0,0)$ at the points $(x_n,0)$ where
\begin{align}
    x_n(t) = \pm\exp\left(\frac{\pi n}{s}-\mathcal{E}t\right)
\end{align}
with $n$-integer. This is consistent with the solutions \eqref{eq:omega1} and \eqref{eq:tilt_omega} for which the level sets $\omega=0$ would be the analogue of the unstable manifold. The parameter $r$ describes the tilt of the unstable manifold at the intersection points as illustrated in Figure \ref{fig:fractal_soln}(a).

Let us assume the adjacent hyperbolic region is centered at the saddle $(-2\ell_0,0)$ whose unstable manifold it tangent to the $x$-axis. Near that saddle, 
the vorticity field is given by \eqref{eq:tilt_omega}, rotated by $90^\circ$ to account for the different orientation of the straining flow. Since the logarithmic term in the argument of the cosine dominates, we find 
\begin{align} 
    &\omega \approx \cos(s\ln|\eta|)=\cos(s\ln|q/\chi|+\phi_0),
\end{align}
where $\phi_0=2\ln(h/\ell_0)$ is a constant which can be absorbed into the definition of the origin of time. Imposing the hyperbolic envelope, we have
\begin{align} \label{eq:manf_soln}
    \omega = \cos(s\ln|q/\chi|) e^{-q^2},
\end{align}
A representative vorticity field described by this solution is shown in Figure \ref{fig:fractal_soln}(b). The corresponding enstrophy spectrum is shown in Figure \ref{fig:fractal_soln}(c) and has the form of a power-law, but now with a non-integral exponent $\alpha$, reflecting the fractal nature of the corresponding vorticity field. Indeed, the large-scale flow considered here is analogous to the baker's map \citep{arnold1968} and the corresponding small-scale vorticity field is analogous to the fractal structures generated by the baker's map.

\begin{figure}
    \centering
    \subfigure[]{\includegraphics[height=0.3\textwidth]{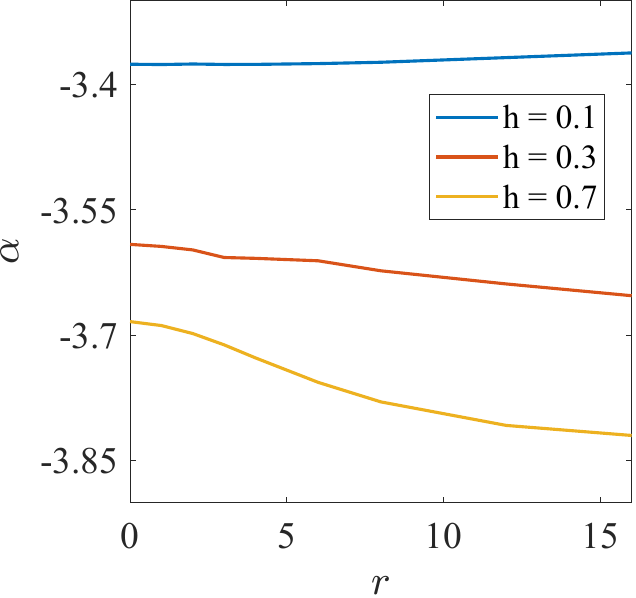}}\hspace{1mm}
    \subfigure[]{\includegraphics[height=0.3\textwidth]{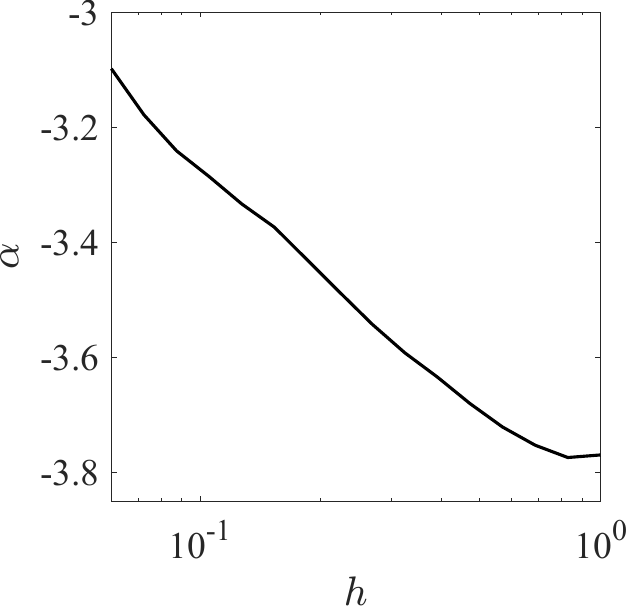}}
\hspace{1mm}
    \subfigure[]{\includegraphics[height=0.3\textwidth]{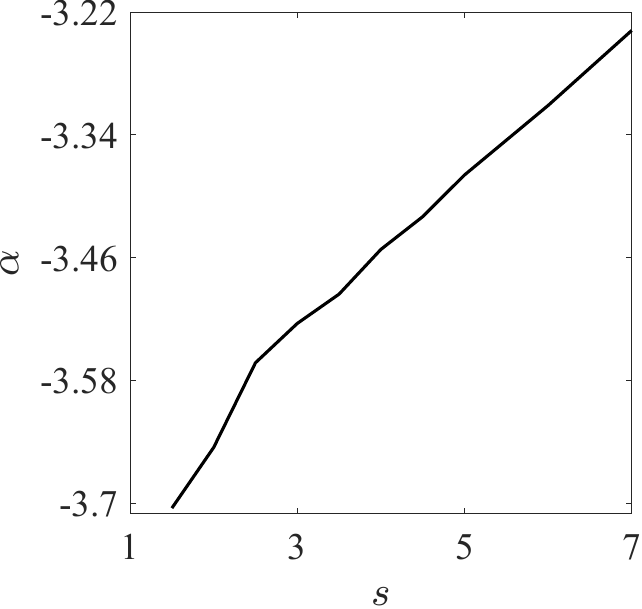}}
    \caption{Scaling exponent $\alpha$ of the enstrophy spectrum $H\propto k^{\alpha+2}$ for the vorticity field \eqref{eq:manf_soln}. The scaling exponent as a function of $r$ for $s = 3$, evaluated at different values of $h$ (a). The scaling exponent as a function of the hyperbolic region size $h$ for $r = 10$, and $s = 4$ (b). The scaling exponent as a function of the strain rate $s$, where $h = 0.2$, and $r = 10$ (c).}
    \label{fig:fractal_exp}
\end{figure}

To confirm that this finding is not an artifact of a particular choice of parameters, we have computed the enstrophy spectrum for a range of parameters and verified that it retains the power-law scaling. The dependence of the scaling exponent on several of the parameters is shown in \autoref{fig:fractal_exp}. For instance, changing the tilt of the unstable manifold (and hence the filaments' orientations) has a relatively weak effect on the value of $\alpha$ and can lead to both an increase and a decrease in the exponent, depending on the thickness of the hyperbolic region, as illustrated in \autoref{fig:fractal_exp}(a). 
On the other hand, the exponent was found to depend rather sensitively on the thickness of the hyperbolic region, decreasing with $h$, as shown in \autoref{fig:fractal_exp}(b). 
The scaling exponent was also found to depend on the frequency of the time-periodic component of the large-scale flow, increasing with $s$ as shown in \autoref{fig:fractal_exp}(c). 
While the characteristic values of $s$ in DNS considered here are found to be smaller than unity, we could not compute the Fourier spectrum of \eqref{eq:manf_soln} with accuracy sufficient to reliably determine the scaling exponent for $s\lesssim 1$
due to the limited resolution of the computational grid (which is constrained by the amount of available memory). As \autoref{fig:fractal_soln}(c) illustrates, even for $s=3$, the inertial range corresponds to rather high wavenumbers. Indeed, given the fractal nature of this vorticity field, very fine meshes are required to properly resolve all the relevant length scales. Note however that the solution \eqref{eq:manf_soln} is meant to illustrate the importance of the interaction effects qualitatively, not reproduce numerical solutions for the vorticity field and their spectra quantitatively.

\section{Discussion}\label{sec:discussion}

The useful analogy with advection of passive scalars \citep{kraichnan1975,kraichnan1971} yields substantial insight into the physical mechanisms of the direct cascade. Previous numerical simulations suggested that this analogy holds in the hyperbolic regions of the flow \citep{lapeyre2001}. The analysis presented in Section \ref{sec:selfsim} shows that this analogy is much more nuanced. Specifically, we find the small-scale vorticity to behave as a passive scalar in regions where (i) the vorticity of the large-scale flow is nearly constant and (ii) the curvature of the vorticity filaments representing small scales is low. In particular, even inside the hyperbolic regions, the sharp folds of the vorticity filaments are not simply advected by the large-scale flow ${\bf u}_l$; the small-scale flow ${\bf u}_s$ also plays an important role. On the other hand, we may find this analogy to also hold inside the elliptic regions of the flow. For instance, for the stationary low-frequency forcing considered here, large-scale vorticity inside vortex cores is often found to be relatively uniform -- the flow field shown in Figure \ref{fig:flow}(a) represents an example of this -- so that small-scale vorticity there satisfies the same evolution equation as in the hyperbolic regions and can also be considered a passive scalar.

\begin{figure}
    \centering
    {\includegraphics[width=0.8\textwidth]{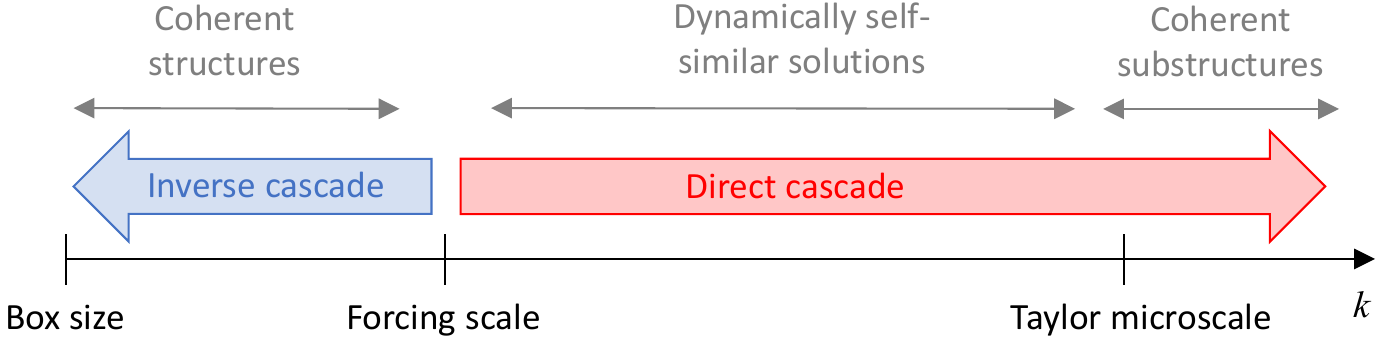}}
    \caption{The relationship between the spatial scales, the cascades, and various classes of exact solutions of the governing equations in high-$Re$ two-dimensional turbulence.}
    \label{fig:scales}
\end{figure}

The physical mechanism of the direct cascade described in this paper provides an explicit, quantitative relation between the properties of coherent structures and the scaling of enstrophy and energy in the inertial range. This mechanism refines the qualitative picture proposed by Kraichnan which involves stretching of filamentary vorticity in the straining regions of the large-scale flow. As shown here, the dynamics of vorticity filaments are described by a novel class of exact, self-similar (and hence scale-free) solutions of the Euler equation, naturally leading to a power-law scaling. Unlike the solutions of the Navier-Stokes equation describing {\it coherent substructures}  \citep{deguchi2015,eckhardt2018,yang2019,doohan2019,azimi2020} or the solutions of the Euler equation describing large-scale {\it coherent structures} \citep{zhigunov2022}, the dynamically self-similar solutions reported here span the entire inertial range associated with the direct cascade. The relationship between different classes of solutions and the two cascades in high-$Re$ two-dimensional turbulence is summarized in Figure \ref{fig:scales}. It is worth emphasizing that, just as in the case of exact solutions of the Euler equation describing large-scale flows \citep{zhigunov2022}, dynamically self-similar solutions describing the direct cascade belong to families spanned by an infinite number of continuous parameters. 


Vortex thinning in the hyperbolic regions of the large-scale flow is exponentially fast. In contrast, the mechanism considered by \citet{moffatt1986} and \citet{gilbert1988} describes a relatively slow linear stretching of vorticity filaments in the vortical (elliptic) regions. Therefore, while both mechanisms operate at the same time, the former plays the dominant role in the direct cascade and explains how power-law spectra can emerge on rather short time scales. Of course, the presence of these two mechanisms does not exclude other physical mechanisms that may contribute to the transport of enstrophy between scales. For instance, vorticity filaments can be stretched exponentially fast inside the vortex cores. Indeed, a time-dependent large-scale flow is generally expected to generate chaotic advection and global mixing in such circular geometries, provided various symmetries are broken \citep{grigoriev2005}. We may also find exponential stretching of vorticity filaments at the edges of vortex cores, where the frequency of the time-dependent component of the flow is resonant with the frequency at which fluid elements are advected around the vortex by the time-independent component, as predicted by the KAM theory \citep{cartwright1999}. Indeed, pronounced vortex filaments are found in these narrow regions as well, as illustrated by Figure \ref{fig:flow}(a).

Time-dependence of the large-scale flow, which has not been fully appreciated previously, plays a critical role in the evolution of vortex filaments in the hyperbolic regions. Chaotic advection generally requires both stretching and folding. While stretching is generic in any flow with a nontrivial structure, folding is a direct result of time dependence. It should be emphasized that the mechanism described by \citet{moffatt1986} and \citet{gilbert1988} involves no time-dependence and hence no folding. In the hyperbolic regions, the structure of vorticity filaments, and the associated Fourier spectrum, is crucially impacted by both the time-dependent and time-independent component of the large-scale flow. In particular, we find the scaling exponent $\alpha$ to depend on the ratio $s$ of the respective time scales as well as other parameters such as $\ell_0$, $h$ and $r$ which also depend on both components.
Note that the DNS considered here feature large-scale flows that are nearly time-periodic on relatively short time scales \citep{zhigunov2022}, with slowly drifting parameters $s$, $\ell_0$, $h$ and $r$. 
Since the stretching of filaments is exponentially fast, the power-law spectrum is established quickly, and the slow parameter drift leads to a correspondingly slow variation of the exponent $\alpha$ shown in Figure \ref{fig:spectrum}(b).

Our results suggest that, for the direct cascade in two-dimensional turbulence, the presence of fractal structures in the vorticity field (or, equivalently, a fractal dimension of level sets of the vorticity field) implies a fractal value of the scaling exponent $\alpha$ and vice versa. The key observation, however, is not that $\alpha$ is fractal but rather that it is {\it non-universal} and takes {\it irrational} values in some interval which depends on the properties of the associated large-scale flow. Indeed, fractal scaling exponents are hardly special. For instance, under appropriate conditions, one finds a {\it rational} exponent $\alpha=-5/3$ for both the direct cascade in three dimensions and for the inverse cascade in two dimensions even in the absence of fractal flow structure. In both cases, the rational value of $\alpha$ is a direct consequence of two assumptions: (1) that the flow structure is scale-invariant and (2) that $E(k)$ depends only on one dimensional parameter (the energy dissipation rate $\varepsilon_\alpha$), with dimensional analysis requiring $E(k)\propto \varepsilon_\alpha^{2/3}k^{-5/3}$ \citep{boffetta2012}. For the direct cascade in two dimensions, similar assumptions leads to the classical KLB prediction $E(k)\propto \eta_\nu^{2/3}k^{-3}$, where $\eta_\nu$ is the enstrophy dissipation rate. The presence of large-scale coherent structure introduces four additional dimensional parameters: $\mathcal{E}$, $\Omega$, $\ell_0$, and $h$. On dimensional grounds, we find $E(k)\propto \eta_\nu^{2/3}k^{-3}f(s,h/\ell_0,k\ell_0)$ in the inertial range, where $f$ is an arbitrary function. This functional form places no restrictions whatsoever on the scaling exponent $\alpha$ and, in particular, allows it to take irrational values.

The KLB theory crucially relies on the assumption that stretching of small-scale vorticity patches in different hyperbolic regions is uncorrelated. Deviations from the KLB predictions are expected whenever this assumption breaks down. Indeed, for the numerical simulations of turbulence considered here, the large-scale flow, a representative snapshot of which is shown in Figure \ref{fig:flow}, has two hyperbolic regions. The dynamics of small-scale vorticity in the two hyperbolic regions are strongly correlated. In particular, the stretching (contracting) direction in one hyperbolic region is well-aligned with the contracting (stretching) direction in the other. To understand and describe the non-universal aspects of the direct cascade and, in particular, determine the enstrophy/energy scaling in the inertial range, these correlations imposed by the large-scale flow have to be properly accounted for. 
A key consequence of these correlations is the pronounced filamentary structures present in the vorticity field, clearly seen in \autoref{fig:flow}, at the entrances to both hyperbolic regions.
As we have shown in Section \ref{sec:fractal}, it is the presence of these vorticity filaments with a broad spectrum of spatial frequencies that is responsible for the emergence of self-similar structure of small-scale vorticity inside the hyperbolic regions characterized by a fractal scaling exponent.

As Figure \ref{fig:flow} illustrates, at least for the forcing considered here, vorticity filaments are aligned along the contracting direction of the large-scale flow, at the entrances to both hyperbolic regions. 
The dynamics of vorticity filaments with such orientation is described, until that orientation changes, by the self-similar solution \eqref{eq:self-sim-} representing the flux of enstrophy towards large, rather than small, scales (backscatter). A proper quantitative description of backscatter is crucial for subgrid-scale modeling. However, correlations induced by the large-scale flow are accounted for neither in the KLB theory nor in the stochastic approaches to modeling eddy viscosity \citep{kraichnan1976,leslie1979,leith1990}. Backscatter amplifies the large scales at the expense of the small scales and should therefore make the spectra of enstrophy and energy steeper, in agreement with the results of numerical simulations. It is worth emphasizing that enstrophy flows in both directions at all times, with the flux towards small (large) scales in regions where small-scale vorticity filaments are preferentially aligned along the expanding (contracting) direction of the large-scale flow. Backscatter, and consequently the energy/enstrophy scaling in the inertial range, are non-universal and cannot be described properly by statistical descriptions agnostic of coherent structures.

It is natural to ask, how much of the analysis presented here would carry over to a more general situation where the large-scale flow is aperiodic. 
The flow domain can still be decomposed into elliptic and hyperbolic regions \citep{haller2001}.
Even though the topology of the large-scale flow will evolve in this case, i.e., some hyperbolic regions will be destroyed, others will emerge, and the ``neighbors'' will change, in the presence of coherent structures, these topological changes are slow while advection and stretching of vorticity filaments is fast (the characteristic time scale for the latter processes is $\tau=\mathcal{E}^{-1}$). This implies that a given vorticity filament will only return to the same hyperbolic region a finite number of times, rather than an infinite number of times as in the time-periodic case. Recall however that the thickness of a filament decreases exponentially quickly; its thickness becomes comparable to the Taylor microscale $\ell_t$ on the short time scale $\tau\ln(k_f/k_t)\sim\ln(Re)/(2\mathcal{E})$ at which point that filament is destroyed by viscosity, whether the flow is time-periodic or not. Hence, for a finite $Re$, the emergence of fractal structures and scaling with a fractal exponent in the inertial range only requires a very small number of returns and time-periodicity of the large-scale flow is not essential. In fact, a vorticity filament may travel through a sequence of different hyperbolic regions and vanish before ever returning to where it was born. The essential feature of the large-scale flow is {\it time-dependence} causing folding of the filaments which, together with stretching, leads to chaotic mixing, even if folding is not time-periodic. Indeed, by Poincare-Birkhof theorem, a {\it steady} two-dimensional flow does not generate Lagrangian chaos and hence cannot produce fractal structures. In the absence of coherent structures, hyperbolic regions are themselves small and the topology of the flow changes quickly, destroying the long-term correlations between neighboring hyperbolic regions that play a key role in generating fractal structures.


\section{Conclusion}\label{sec:conclusions}
We introduced here a mechanistic, dynamical description of the direct cascade in two-dimensional turbulence in the presence of  coherent structures. This description involves two qualitatively different classes of solutions of the Euler equation: (weakly unstable) time-periodic solutions describing the large scales, i.e., vortex condensate, and dynamically self-similar solutions describing the small scales or, more precisely, the scales corresponding to the inertial range. The self-similar, i.e., scale-free, nature of the latter class of solutions is the fundamental reason why power-law scaling of energy and enstrophy is found in the inertial range, regardless of whether the scaling exponent is integral or fractal.

In the absence of correlations between different hyperbolic regions of the large-scale flow, our description reproduces the scaling predicted by the KLB theory. In particular, we find that enstrophy flows strictly from large to small scales with no backscatter. In the presence of pronounced coherent structures, adjacent hyperbolic regions of the large-scale flow become strongly correlated leading to a complete breakdown of the KLB theory. These correlations have a dramatic effect on the direct cascade. In particular, they lead to a substantial and persistent backscatter, with the enstrophy flowing from large to small scales at the exit(s) and from small to large scales at the entrance(s) of the hyperbolic regions, with the thickness of a typical filament repeatedly growing and shrinking at it is advected by the large-scale flow. Backscatter leads to steeper spectra in the inertial range than what KLB theory predicts. In fact, it is not uncommon to find the scaling exponent $\alpha$ as high as $-5$ in the DNS considered here. Other numerical and experimental studies of forced turbulence reported different exponents, but the common finding is that $\alpha<-3$. The entirety of available results clearly demonstrates that, in the presence of coherent structures, the scaling in the inertial range is nonuniversal and the value of the exponent depends on the properties of the large-scale flow.

Correlations are also responsible for the scaling exponent taking fractal values which reflect the emergence of self-similar fractal structures in the vorticity field. Such fractal structures represent ``stacks'' of vorticity filaments of very different thickness generated by repeated stretching and folding of small-scale vorticity by the large-scale flow. Their properties -- their orientation and shape, their  fractal dimension and the scaling exponent -- are controlled by both the large-scale flow -- namely the thickness and size of the hyperbolic regions, the ratio of its temporal frequency to the local strain rate, and the angle at which the stable and unstable manifolds of adjacent saddles intersect -- and the forcing which injects both energy and enstrophy into the flow. In particular, for large-scale flows that are time-periodic, these fractal structures are described by solutions that are self-similar in space and periodic in time.

In conclusion, let us point out some limitations of the present work. First of all, while the large-scale flow, {\it for the steady, low-frequency forcing considered here,} tends to be nearly time-periodic for extended periods of time, this property is unlikely to be representative of turbulent flows driven by other types of forcing (e.g., stochastic-in-time or high-frequency in space). Corresponding large-scale flows can be nearly steady or even completely aperiodic. While, for such flows, one may still be able to consider small-scale vorticity to be effectively advected by the large-scale flow, this description may break down in regions where the large-scale vorticity itself exhibits fast variation (e.g., in the presence of internal boundary layers). In the aperiodic case, the topology of the large-scale flow will also evolve, presenting additional complications. Furthermore, even in the regions where the large-scale vorticity is nearly constant, small-scale vorticity cannot be considered a passive scalar where vorticity filaments have a high curvature. These limitations imply that the dynamical description presented here requires further refinement and validation before it can be extended to other turbulent flows featuring different coherent structures.

\medskip\noindent
{\bf Acknowledgements.} This material is based upon work supported by the National Science Foundation under grant no. 2032657.

\medskip\noindent
{\bf Declaration of interests.} The authors report no conflict of interest.

\appendix

\section{Effect of viscosity}\label{app:A}


In the presence of viscosity, self-similarity of the small-scale vorticity field in a hyperbolic region will break down when the filament thickness becomes comparable to the Taylor microscale $\ell_t$. 
Equation \eqref{eq:vort1} should be replaced with
\begin{align}\label{eq:vort-visc}
    \partial_t\omega-\mathcal{E}x\partial_x\omega-\nu\partial_x^2\omega=0.
\end{align}
Small viscosity implies that the ratio $\epsilon=\nu/\mathcal{E}=\ell_t^2$ is small compared with $\ell_0^2$. In the inertial range, the product $\epsilon x^{-2}=(\ell_t/x)^2\ll 1$, and we can represent the effect of viscosity perturbatively using a WKB expansion
\begin{align}\label{eq:WKB}
    \omega= \frac{1}{2}\sum_{n=1}^{\infty}\bar{a}_n^\pm \exp\left[\Theta_{n0}(\chi)+\epsilon x^{-2}\Theta_{n1}(\chi)+\epsilon^2x^{-4}\Theta_{n2}(\chi)+\cdots\right]+\text{c.c.},
\end{align}
where c.c. denotes complex conjugate.
For $\epsilon=0$, the corrections disappear and we should recover the inviscid solution \eqref{eq:invis_soln}, hence
\begin{align}
\Theta_{n0}(\chi)=ins\ln|\chi|+i\bar{\phi}^\pm_n=ins\ln|x/\ell_0|+in\Omega t+i\bar{\phi}^\pm_n
\end{align} 

Substituting \eqref{eq:WKB} into \eqref{eq:vort-visc} and collecting terms of different orders in $\epsilon$, we find 
\begin{align}\label{eq:s1_general}
    \Theta_{n1}(\chi)&=\frac{\Theta_{n0}''(\chi)+\Theta_{n0}'(\chi)[\Theta_{n0}'(\chi)-1]}{2},
\end{align}
etc. Substituting the expression for $\Theta_{n0}(\chi)$ into this equation, we find
\begin{align}\label{eq:s1}
    \Theta_{n1}&=\frac{-ins-n^2s^2}{2}.
\end{align}
The vorticity is therefore given by the spatially modulated version of the original self-similar solution
\begin{align}
    \omega(x,t) \approx \sum \limits_{n=1}^{\infty} a_n^\pm(x)\cos\left[n s \ln|x/\ell_0| + n \Omega t+\phi_n^\pm(x)\right],
\end{align}
where, to leading order in $\ell_t/x$,
\begin{align}
    \phi_n^\pm(x)=\bar{\phi}^\pm_n-\frac{ns}{2}\frac{\ell_t^2}{x^2}
\end{align}
and
\begin{align}\label{eq:amp}
    a_n^\pm(x)=\bar{a}_n^\pm\exp\left[-\frac{n^2s^2}{2}\frac{\ell_t^2}{x^2}\right].
\end{align}
Note that, for $|x|\gg s\ell_t$, the amplitudes $a_n^\pm(x)\approx \bar{a}_n^\pm$ are essentially constant and $\phi_n^\pm(x)\approx\bar{\phi}^\pm_n$ while, for
\begin{align}\label{eq:diss_rng}
|x|<\sigma\ell_t,
\end{align}
the amplitudes $a_n^\pm(x)$ become exponentially small. It is straightforward to compute the solutions to a higher order in perturbation theory although additional corrections do not change any results presented here qualitatively.

\section{Fourier spectrum of the vorticity field}\label{app:B}

In this Appendix we derive the power spectra of analytical solutions which represent the enstrophy flux from large to small scales in non-interacting hyperbolic regions. We discuss the limits of thin (small-$h$) and thick (large-$h$) hyperbolic regions as well as the effect of varying the angle between the expanding and the contracting direction.

\subsection{Thin hyperbolic regions}\label{app:B1}
 
\begin{figure}
    \centering
    \subfigure[]{\includegraphics[height=0.36\columnwidth]{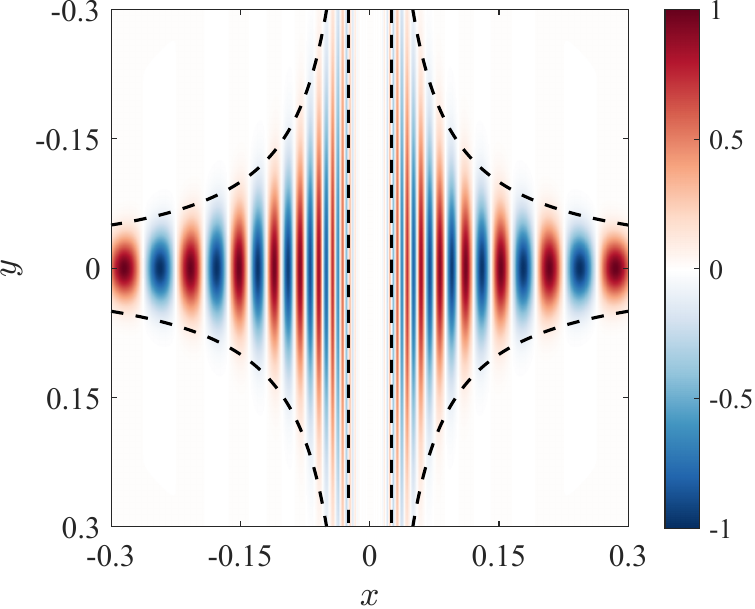}}
    \hspace{1mm}
    \subfigure[]{\includegraphics[height=0.36 \columnwidth]{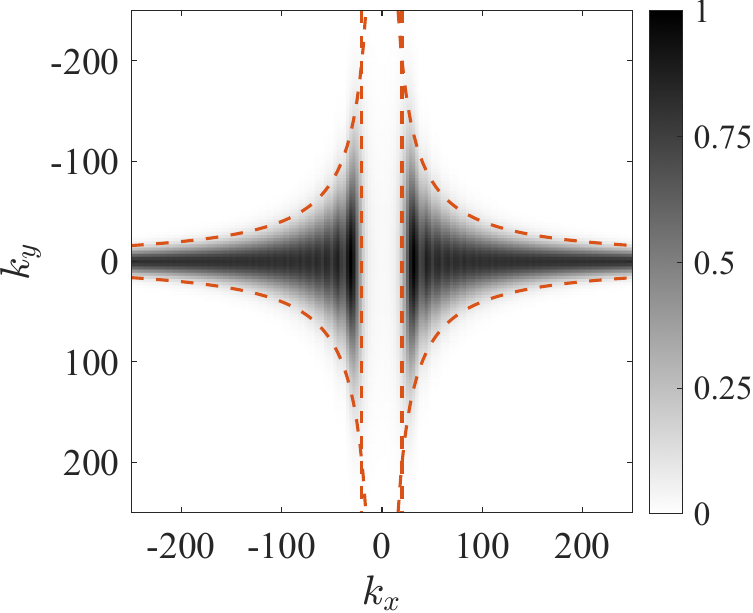}}
    \caption{A snapshot of the vorticity field $\omega({\bf x},t)$ defined by Eq. \eqref{eq:omega1} in the physical space for $s=20$, $h=0.1$, $\ell_0=1$, and $Re=10^5$ (a). The black dashed line represents the boundaries described by Equations \eqref{eq:xy} and \eqref{eq:diss_rng}. 
    The corresponding (normalized) power spectrum $\hat{H}({\bf k})$ (b). The boundaries defined by Equations \eqref{eq:ky} and \eqref{eq:ks} are shown as red dashed lines.}
    \label{fig:2Dspec}
\end{figure}
 
A representative snapshot of the vorticity field defined by Equations \eqref{eq:visc_soln} and \eqref{eq:amp_hyp} is shown in Figure \ref{fig:2Dspec}(a). The dashed lines represent the boundaries of the spatial region contributing to the inertial range. In particular, we see that vorticity field is strongly attenuated in the region $|x|<s\ell_t$ due to the action of viscosity, as discussed in Appendix \ref{app:A}. 
The Fourier spectrum of this vorticity field is given by
\begin{align}\label{eq:2Dint}
    \hat{\omega}({\bf k},t)&\approx 
    \sum \limits_{n=1}^{\infty} \frac{ \bar{a}_n^-}{2} \int_{-\ell_0}^{0} dx\int_{-\ell_0}^{\ell_0} dy\exp\left[f_{ns}({\bf k},{\bf x}) + in\Omega t +i\bar{\phi}_n^-\right]\nonumber\\
    &+\sum \limits_{n=1}^{\infty} \frac{ \bar{a}_n^+}{2} \int_{0}^{\ell_0} dx\int_{-\ell_0}^{\ell_0} dy \exp\left[f_{ns}({\bf k},{\bf x}) + in\Omega t +i\bar{\phi}_n^+\right]  +\textit{c.c.},
\end{align}
where
\begin{align}
    f_r({\bf k},{\bf x})= ir\ln|x|-\frac{ir+r^2}{2k_t^2 x^2}-\frac{x^2y^2}{h^4}+i{\bf k}\cdot{\bf x},
\end{align}
and the dependence on $\ell_0$ is absorbed in the definition of the phase $\bar{\phi}_n^\pm$.

The integrals can be evaluated using the saddle-point approximation. 
The saddle point ${\bf x}_s$ is found by solving the equation $\nabla f_r({\bf k},{\bf x}_s)=0$, 
\begin{align}\label{eq:saddle}
    x_s &= - \frac{r}{k_x} X(\beta_r), \nonumber\\ 
    y_s &=  \frac{i k_yh^4}{2x_s^2}.
\end{align}
Here $X(\beta)$ is the dominant root of the cubic equation $X^2 - X^3 +\beta=0$ which corresponds to the limit $X(0)=1$,
\begin{align}
    \beta_r = 
    -i \frac{\rho}{2 r^3}+\frac{1-ir}{r^2}\kappa^2,
    \end{align}
and we introduced the short-hand notations $\kappa=k_x/k_t$ and $\rho = k_x^2 k_y^2 h^4$. 

So long as $\left|\textit{Re}[x_s]\right|<\ell_0$, the saddle point approximation yields 
\begin{align}\label{eq:omegahat}
    \hat{\omega}({\bf k},t)&\approx 
    \sum \limits_{n=1}^{\infty} \frac{\pi \bar{a}_n^- \exp\left[f_{ns}({\bf k},{\bf x}_s)+in\Omega t+i \bar{\phi}_n^-\right]}{2\sqrt{-g_{ns}({\bf k},{\bf x}_s)}} \nonumber\\
    &+\sum \limits_{n=1}^{\infty} \frac{\pi \bar{a}_n^+ \exp\left[f_{ns}({\bf k},{\bf x}_s)+in\Omega t+i \bar{\phi}_n^+\right]}{2\sqrt{-g_{ns}({\bf k},{\bf x}_s)}}
    +\textit{c.c.}
\end{align}
where
\begin{align}
  f_r({\bf k},{\bf x}_s)& = 
  ir\ln\left|\frac{r}{k_x}\right|+ir F(\beta_r),\nonumber\\
  F(\beta)&=\ln X(\beta) - X(\beta)-\frac{\beta}{2X^2(\beta)}
\end{align}
and
\begin{align}
    g_r({\bf k},{\bf x}_s)&=\det(\nabla\nabla f_r)|_{{\bf x}_s} = \frac{ir G(\beta_r)}{h^4},\nonumber\\
  G(\beta)&=2 + 6\frac{\beta}{X^2(\beta)}.
\end{align}

Let us define the average of the enstrophy spectrum over a period of the large-scale flow
\begin{align}\label{eq:Hhat}
    \hat{H}(\mathbf{k}) = \frac{1}{T}\int_0^{T} |\hat{\omega}({\bf k},t)|^2 dt.
\end{align}
Using appropriate orthogonality relations, we find
%
\begin{align}\label{eq:2Dspec}
    \hat{H}(\mathbf{k})\approx
    2\pi^2h^4\sum_{n=1}^{\infty} H_n A(n\sigma,\rho,\kappa),
\end{align}
where
\begin{align}\label{eq:spec_amp}
    H_n = \frac14\left|\bar{a}_n^-e^{i\bar{\phi}_n^-}+\bar{a}_n^+e^{i\bar{\phi}_n^+}\right|^2
\end{align}
and the wavenumber dependence is contained entirely in the amplitudes \begin{align}\label{eq:Aexact}
A(r,\rho,\kappa)= 
\frac{e^{\text{Re}[2f_r]}}{r|G(\beta_r)|}.
\end{align}
%

\begin{figure}
    \centering
     \subfigure[]{\includegraphics[width=0.45\textwidth]{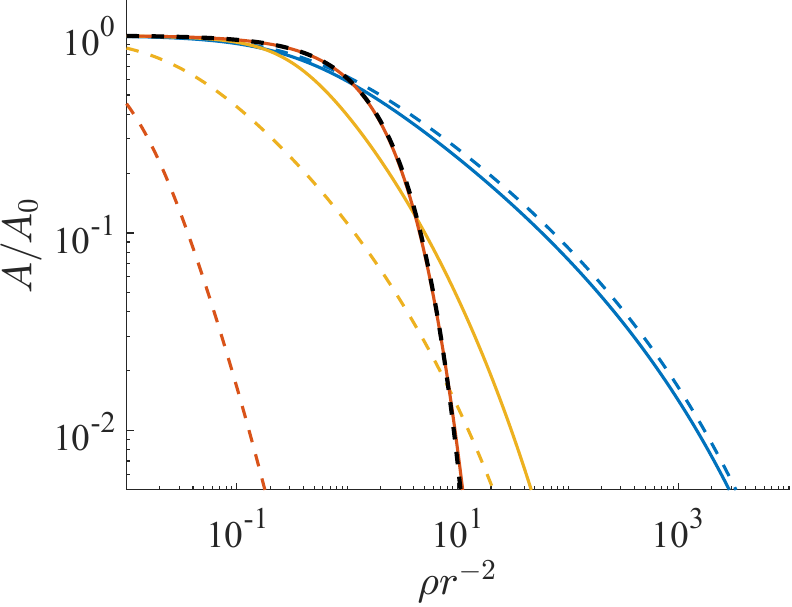}}
     \hspace{1mm}
     \subfigure[]{\includegraphics[width=0.45\textwidth]{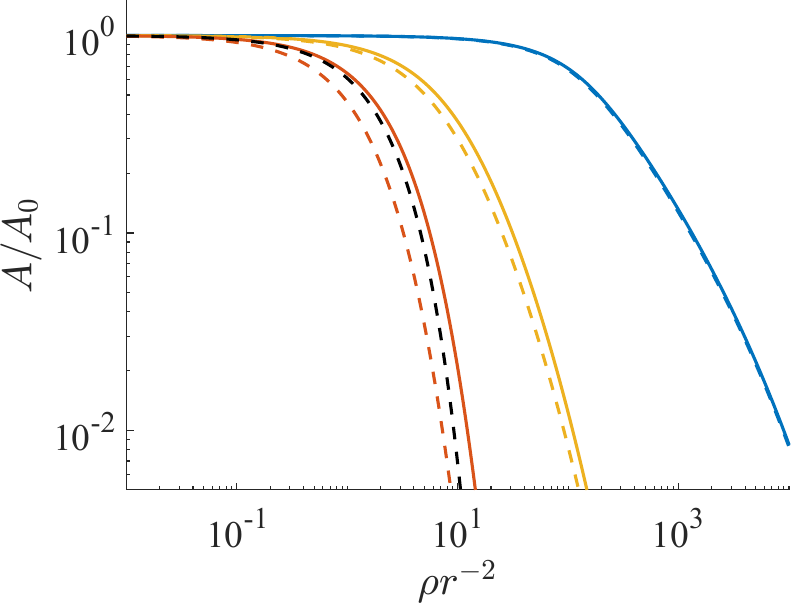}}
    \caption{The amplitude function $A(r,\rho,\kappa)$, normalized by $A_0=A(r,0,\kappa)$, for $\kappa=0.1$ (a) and $\kappa=2$ (b) with $r=20$ (red), $r=1$ (yellow), and $r=0.05$ (blue). The results based on the exact solution $X(\beta)$ (solid lines) are compared with those based on the power series expansion of $X(\beta)$ in the limits of low $\beta$ (black dashed line) and high $\beta$ (colored dashed lines).}
    \label{fig:A}
\end{figure}

Note that, in the absence of viscosity, the enstrophy spectrum depends on the wavenumber only through the nondimensional combination $\rho$, with the amplitude $A(r,\rho,\kappa)$ being a decreasing function of $\rho$. Therefore, at high $Re$, the enstrophy is confined to a hyperbolic region in the Fourier space as well, as illustrated in Figure \ref{fig:2Dspec}(b).
The width of the hyperbolic region can be easily estimated in various limiting cases. In the limit of $r$-large, $\beta$ is small and we can use the Taylor expansion
\begin{align}\label{eq:X}
    X(\beta)=1+\beta-2\beta^2+\cdots.
\end{align}
Keeping the leading order terms in $\beta$, 
we find
an explicit expression for the amplitude
\begin{align}\label{eq:Ahigh}
  A(r,\rho,\kappa)\approx
  \frac{1}{2r}\exp\left[-\frac{\rho}{2r^2}-\kappa^2\right].
\end{align}
The approximation \eqref{eq:X} is accurate for $r\gtrsim 10$, but it starts to break down for smaller values of $r$, especially when $\kappa$ is low. This is illustrated in Figure \ref{fig:A} which compares this approximation with the exact result \eqref{eq:Aexact} for $\kappa=0.1$ and $\kappa=2$ representative of the inertial and dissipation range, respectively. 
For $r$-small, one should instead consider the limit of $\beta$-large where
\begin{align}
    X(\beta)=\beta^{1/3}+\frac{1}{3}
    +\frac{1}{9}\beta^{-1/3}+\cdots
\end{align}
and, to leading order in $\beta$, 
\begin{align}\label{eq:Alow}
  A(r,\rho,\kappa)\approx
  \frac{1}{6r\left|\beta_{r}^{1/3}\right|}\exp\left[-3r\text{Re}\left[i\beta_{r}^{1/3}\right]\right].
\end{align}
This approximation is found to be accurate for $r\lesssim 0.1$, as illustrated in Figure \ref{fig:A}.

Let $\bar{n}$ be the dominant harmonic in the Fourier series \eqref{eq:invis_soln} and $\bar{r}\equiv\bar{n}s$. In the inertial range, the amplitude \eqref{eq:Alow} becomes independent of $r$, hence the hyperbolic region is defined by the inequality 
\begin{align}\label{eq:ky}
    |k_xk_y|\lesssim \frac{z(\bar{r})}{h^2},\qquad z(\bar{r})=
    \begin{cases}
        1,& \bar{r}\lesssim 0.1, \\
        2\bar{r}, & \bar{r}\gtrsim 10.
    \end{cases}
\end{align}
Furthermore, the saddle-point approximation which underlies Equation \eqref{eq:Aexact} is only accurate when the saddle lies inside the integration domain, i.e., $|x_s|<\ell_0$ which corresponds to
\begin{align}\label{eq:ks} 
    |k_x|\gtrsim k_s\equiv\frac{\bar{r}}{\ell_0}.
\end{align}
For $|k_x|<k_s$, the amplitude $A(r,\rho,\kappa)$ becomes exponentially small. As Figure \ref{fig:2Dspec}(b) illustrates, the inequalities \eqref{eq:ky} and $\eqref{eq:ks}$ describe well the boundaries of the region in the Fourier space outside of which the enstrophy spectrum essentially vanishes.
The maximal width of the enstrophy spectrum in the $k_y$ direction can be estimated by setting $k_x=k_s$ in Equation \eqref{eq:ky}, yielding
\begin{align}\label{eq:kh} 
     \max_{k_x}|k_y|\lesssim k_w\equiv\frac{z(\bar{r})\ell_0}{\bar{r}h^2}.
\end{align}

The enstrophy scaling can be determined by integrating $\hat{H}(\mathbf{k})$ over the annular region $k<|{\bf k}|<k+dk$. For $k\gtrsim \max(k_s,k_w)$, the dominant contribution to the integral comes from the two arcs near the $k_x$ axis where the annular region intersects the hyperbolic region defined by the inequality \eqref{eq:ky}. For $k\gtrsim k_a$, where 
\begin{align}\label{eq:ka} 
     k_a\equiv\frac{2\sqrt{2\bar{r}}}{h},
\end{align}
these arcs are well-approximated by straight lines with $k_x=k$, such that 
\begin{align}\label{eq:Hgen}
    H(k)=
    4\pi^2h^2\sum \limits_{n=1}^{\infty} 
    H_n\left[\int_0^\infty A(ns,\rho,\kappa)\frac{d\rho}{\sqrt{\rho}}\right]_{k_x=k}\,k^{-1}.
\end{align}
In the inertial range, we can set $\kappa=0$, so the expression in square brackets becomes $k$-independent and we find $H(k)\propto k^{-1}$, as predicted by the KLB theory.

\begin{figure}
    \centering
    \subfigure[]{\includegraphics[width=0.45\textwidth]{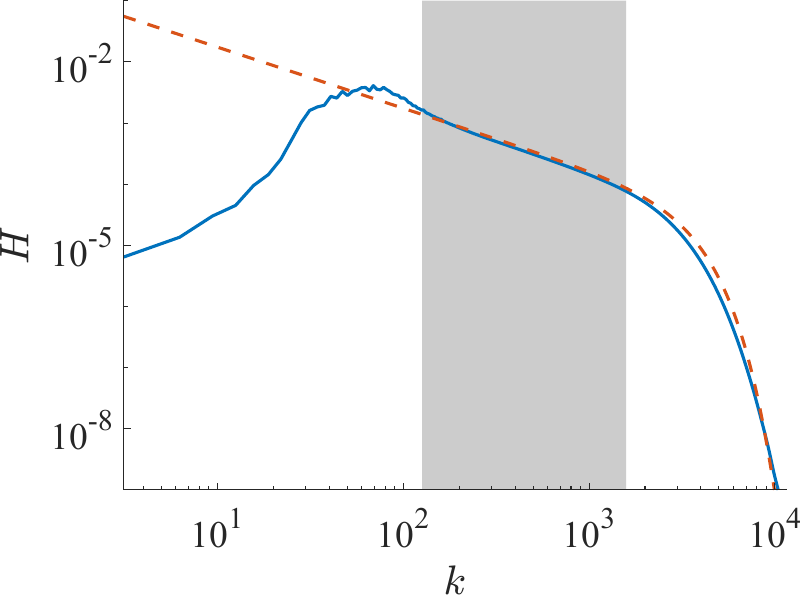}}
    \hspace{1mm}
    \subfigure[]{\includegraphics[width=0.45\textwidth]{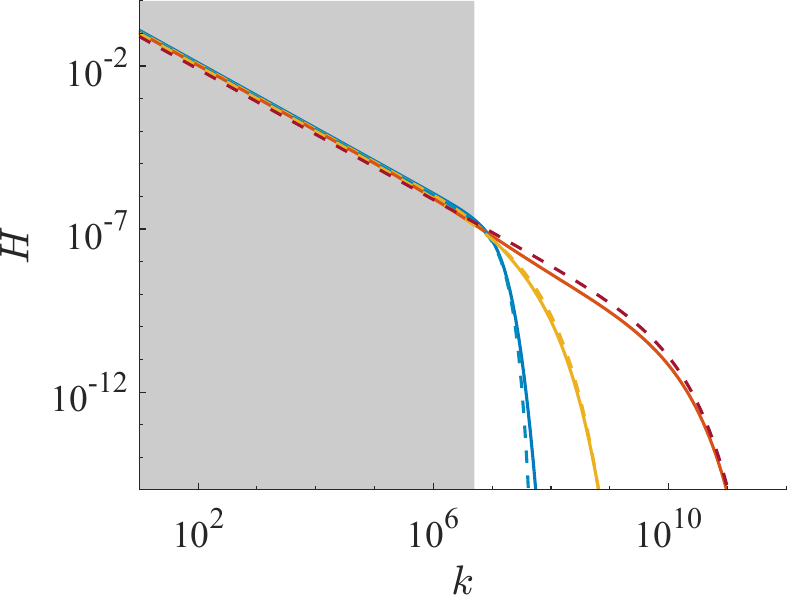}}
    \caption{The enstrophy spectrum $H(k)$. (a) The result obtained by direct numerical evaluation of the time-average \eqref{eq:Hhat}, shown as a solid blue line, is compared with the analytical result \eqref{eq:Hhigh}, shown as a dashed red line, for the vorticity field \eqref{eq:omega1} with $s=20$, $h=0.1$, $\ell_0 = 1$, and $Re = 10^7$. (b) The spectrum calculated using the saddle-point approximation \eqref{eq:Hgen} for $s = 20$ (blue), $s =2$ (yellow), and $s = 0.2$ (red) and $Re = 10^{12}$ (solid lines). Dashed lines show the fits to an expression of the form $H \propto \exp{(-b(k/k_t)^\gamma)} k^{-1}$, where $b=1$ and $\gamma=2$ for $s=20$; $b=0.74$ and $\gamma=0.71$ for $s=1$; $b=0.018$ and $\gamma=2/3$ for $s=0.05$. In both panels, the gray background represents the inertial range $k_c<k<k_d$ whose boundaries are given by Equations \eqref{eq:kd} and \eqref{eq:kc}. 
    }
    \label{fig:enstrophy}
\end{figure}

Further progress can be made in various limiting cases. For $s$-large, the integral in \eqref{eq:Hgen} can be evaluated with the help of \eqref{eq:Ahigh}, yielding
\begin{align}\label{eq:Hhigh}
    H(k) = 2\pi^{5/2} h^2 \sum_{n=1}^{\infty} H_n e^{-(k/k_t)^2}k^{-1}
\end{align}
with the Gaussian envelope describing the effect of viscosity. The corresponding energy spectrum is given by
\begin{align}\label{eq:Ehigh}
    E(k)&\propto k^{-2}H(k)\propto e^{-(k/k_t)^2}\,k^{-3}.
\end{align}
The analytical result \eqref{eq:Hhigh} accurately represents the enstrophy spectrum computed numerically in both the inertial and dissipation range, as Figure \ref{fig:enstrophy}(a) illustrates. 
As expected, the boundary between these two regions is defined by the Taylor microscale $k_t=\ell_t^{-1}=Re^{1/2}\ell_0^{-1}$. Noticeable deviations from power-law scaling are found for $k\gtrsim k_d$, where
\begin{align}\label{eq:kd}
     k_d\equiv\frac{k_t}{2}.
\end{align}
The low-wavenumber boundary of the inertial range should be described by 
\begin{align}\label{eq:kc}
     k_c\equiv\max\left(k_s,k_w,k_a\right).
\end{align}
Both boundaries, $k=k_c$ and $k=k_d$, are indeed in good agreement with the numerically computed spectra corresponding to the self-similar vorticity field \eqref{eq:omega1} as can be seen in Figure \ref{fig:enstrophy}.  
Note that, in the limit of $s\to 0$, $k_c=k_w\propto s^{-1}$, while, in the limit $s\to\infty$, $k_c=k_s\propto s$. Hence, the inertial range should be the widest for intermediate values of $s$.   

For $s$-small, the integrals in \eqref{eq:Hgen} can be evaluated using the approximation \eqref{eq:Alow} for $1\le n\le n_{\text{max}}$, where $n_{\text{max}}\sim 0.1/s$. The integrals are dominated by values of $\rho$ much smaller than unity, and, in this limit, we can use a series expansion
\begin{align}
   3r\beta_r^{1/3}\approx -c_{1,r}\kappa^{2/3} - c_{2,r} \rho\kappa^{-4/3},
\end{align}
where we have defined the coefficients $c_{1,r}= -3(r-ir^2)^{1/3}$ and $c_{2,r}=\frac{i}{2}(r-ir^2)^{-2/3}$. Assuming the series \eqref{eq:invis_soln} truncated at $n_{\text{max}}$ accurately describes the small-scale vorticity field, we can write
\begin{align}
    H(k) \approx \sum_{n=1}^{n_\textit{max}} \int_{0}^{\infty} \frac{2 \pi^2 H_n h^4}{|c_{1,ns}|\kappa^{2/3}} \exp{\left[-\mathrm{Im}(c_{1,ns})\kappa^{2/3} - \mathrm{Im}(c_{2,ns}) \frac{\rho}{\kappa^{4/3}}\right]} \frac{d\rho}{\sqrt{\rho}}.
\end{align}

Evaluating the integrals with the help of the saddle-node approximation, we find
\begin{align}\label{eq:Hlow}
    H(k) \approx \frac{2^{3/2}\pi^{5/2} h^2}{\sqrt{3}}\sum_{n=1}^{n_\textit{max}} H_n e^{-\mathrm{Im}(c_{1,ns})(k/k_t)^{2/3}}k^{-1}, 
\end{align}
where we used the relation $|c_{1,r}|\,[\mathrm{Im}(c_{2,r})]^{1/2}\approx \sqrt{3/2}$ which holds at small values of $r$. It should be pointed out that, for $s$-small and $\max(k_s,k_a)<k<k_w$, the integral of $\hat{H}({\bf k})$ over the annular region should include the contribution from the hyperbolic region near the $k_y$-axis, which would require a correction to Equation \eqref{eq:Hgen}. This correction should be relatively small and therefore is not considered here.  

The accuracy of the approximations \eqref{eq:Hhigh} and \eqref{eq:Hlow} can be confirmed by comparing them with the integral \eqref{eq:Hgen} evaluated numerically. Figure \ref{fig:enstrophy}(b) shows the respective results computed for the vorticity field \eqref{eq:omega1}. For both $s=20$ and $s=0.05$, the enstrophy spectrum is indeed well-described by the analytical results \eqref{eq:Hhigh} and \eqref{eq:Hlow}, respectively. While both approximations break down at the intermediate values of $s$, we can leverage the functional form of these approximations to speculate that the enstrophy spectrum generally has the form
\begin{align}\label{eq:Hint}
     H(k)\propto e^{-b(s)\,(k/k_t)^{\gamma(s)}}\,k^{-1},
\end{align}
where the exponent $\gamma$ interpolates between the values of $2/3$ and 2 for $s$-low and $s$-high, respectively. For the intermediate value $s=1$, we find the spectrum can indeed be fitted quite well by this expression with $b=0.74$ and $\gamma=0.71$.


\subsection{Thick hyperbolic regions}\label{app:B2}

For thick hyperbolic regions, much of the analysis conducted in the limit of $h$-small still applies with minor modifications. In particular, we can take the limit $h\to\infty$ and the $y$ integral in \eqref{eq:2Dint} can be evaluated explicitly, yielding
\begin{align}\label{eq:thin_fourier}
    \hat{\omega}({\bf k},t)&\approx 
    \sum \limits_{n=1}^{\infty} \frac{ \bar{a}_n^-}{2} \int_{-\ell_0}^{0} dx\, \frac{\sin{(k_y \ell_0)}}{\pi k_y} \exp\left[f_{ns}(k_x,x) + in\Omega t +i\bar{\phi}_n^-\right]\nonumber\\
    &+\sum \limits_{n=1}^{\infty} \frac{ \bar{a}_n^+}{2} \int_{0}^{\ell_0} dx\, \frac{\sin{(k_y \ell_0)}}{\pi k_y} \exp\left[f_{ns}(k_x,x) + in\Omega t +i\bar{\phi}_n^+\right]  +\textit{c.c.},
\end{align}
where
\begin{align}
    f_r(k_x,x)= ir\ln|x|-\frac{ir+r^2}{2k_t^2 x^2}+i k_x x.
\end{align}
The $x$ integral can again be evaluated using the saddle point approximation:
\begin{align}\label{eq:omegahat}
    \hat{\omega}({\bf k},t)&\approx 
    \sum \limits_{n=1}^{\infty} \frac{\sqrt{\pi} \bar{a}_n^- \exp\left[f_{ns}(k_x,x_s)+in\Omega t+i \bar{\phi}_n^-\right]}{2\sqrt{-2 f''_{ns}(k_x,x_s)}} \frac{\sin{(k_y \ell_0)}}{\pi k_y} \nonumber\\
    &+\sum \limits_{n=1}^{\infty} \frac{\sqrt{\pi} \bar{a}_n^+ \exp\left[f_{ns}(k_x,x_s)+in\Omega t+i \bar{\phi}_n^+\right]}{2\sqrt{-2 f''_{ns}(k_x,x_s)}} \frac{\sin{(k_y \ell_0)}}{\pi k_y}
    +\textit{c.c.},
\end{align}
where $x_s$ can again be found using \eqref{eq:saddle} with $\rho = 0$. Now we have
\begin{align}
    f_r''(k_x,x_s) &= - \frac{ik_x^2}{r} I(\beta), \nonumber \\
    I(\beta) &= X(\beta)^{-2} + 3 \beta X(\beta)^{-4}.
\end{align}

Computing the radial and temporal averages of $\hat{H}({\bf k},t)$ yields the enstrophy spectrum
\begin{align}
    H(k) \approx \frac{\ell_0}{\sqrt{2\pi}} \sum \limits_{n=1}^{\infty} H_n A(ns,\kappa)\hspace{0.1em}  k^{-2} ,
\end{align}
where the amplitudes $H_n$ remain unchanged and
\begin{align}
  A(r,\kappa) = \frac{r e^{\text{Re}[2f_r]}}{|I(\beta_r)|} .
\end{align}
In the limit of $r$-large, the amplitudes reduce to
\begin{align}\label{eq:A1Dhigh}
  A(r,\kappa)\approx
  \exp\left[-\kappa^2\right]
\end{align}
and in the limit of $r$-small we obtain
\begin{align}\label{eq:A1Dlow}
  A(r,\kappa)\approx
  \frac{r^{1/3}}{3 \kappa^{2/3}} \exp\left[- (\kappa/r)^{2/3} \right],
\end{align}
Note that the latter expression is only valid for $\kappa=O(1)$, and we still find $H(k)\propto k^{-2}$ in the inertial range.


%
%

\subsection{Effect of the angle between expanding and contracting directions}\label{app:B3}

\begin{figure}
    \centering
    \subfigure[]{\includegraphics[width=0.45\textwidth]{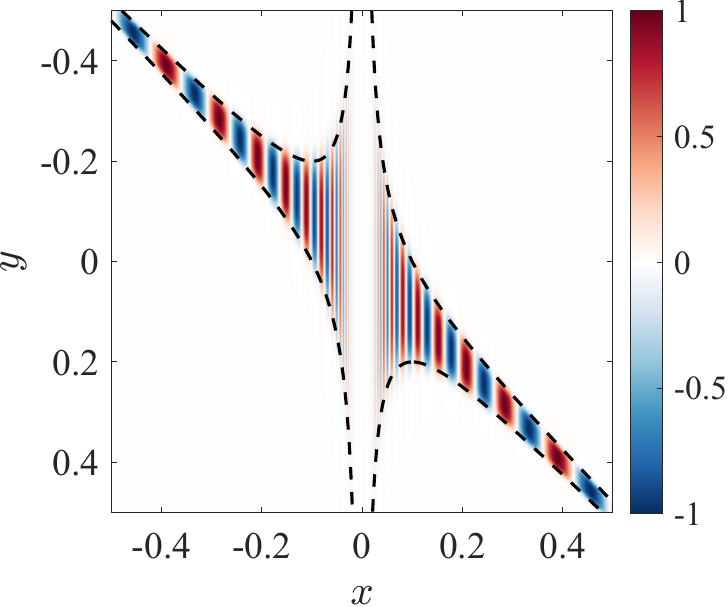}}
    \hspace{1mm}
    \subfigure[]{\includegraphics[width=0.45\textwidth]{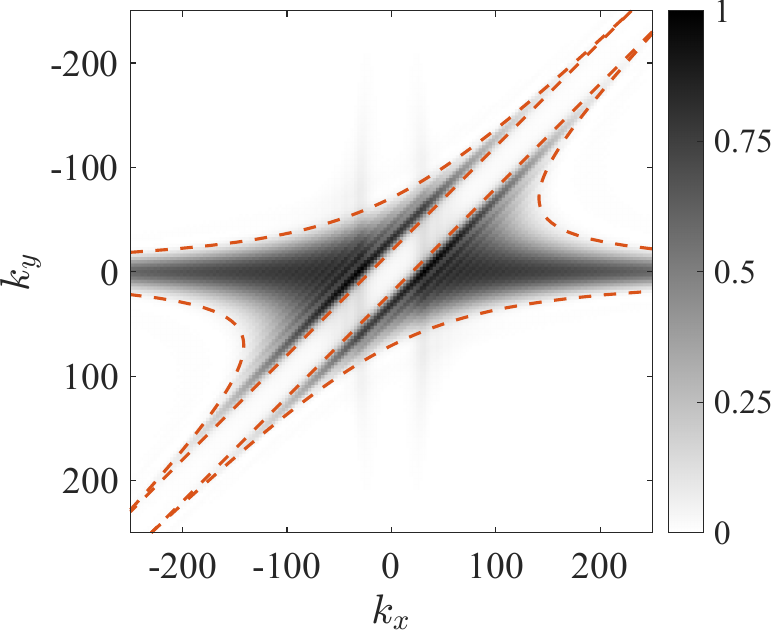}}
    \caption{A snapshot of the vorticity field $\omega({\bf x},t)$ defined by Eq. \eqref{eq:omega1} in the physical space for $s=20$, $h=0.1$, $\ell_0=1$, $Re=10^6$, and $\theta \approx 45^\circ$ (a). The black dashed line represents the boundary of the hyperbolic region given by inequality \eqref{eq:xy_gen}. The corresponding (normalized) power spectrum $\hat{H}(\mathbf{k})$ (b). The boundaries of the hyperbolic region defined by Eqs. \eqref{eq:ky_gen} and \eqref{eq:ks_gen} are shown as red dashed lines.
    }
    \label{fig:2Dspec_gen}
\end{figure}

In the analysis elsewhere in this paper, we assumed the expanding and contracting directions of the large-scale flow to intersect at right angles. However, the key results hold for arbitrary angles $\theta$. Generally, it is convenient to use an orthogonal coordinate system where the $y$ axis is aligned with the expanding direction. For this choice of the coordinate system, small-scale vorticity can still be described by the self-similar solution \eqref{eq:self-sim+}, where $\chi =  |x|e^{\mathcal{E} t}$ and $\eta = |y+x\cot(\theta)|e^{-\mathcal{E}t}$. The shape of the hyperbolic region, 
\begin{align}\label{eq:xy_gen}
    \left|xy+\cot(\theta)x^2\right|\lesssim h^2,
\end{align}
will be change accordingly, as illustrated in \autoref{fig:2Dspec_gen}(a) for $\theta \approx 45^\circ$. 

The enstrophy spectrum will also remain confined to a region of hyperbolic shape in Fourier space, with the ``contracting'' and ``expanding'' directions making an angle $\theta$, as illustrated in \autoref{fig:2Dspec_gen}(b). The boundaries of this region are given by 
\begin{align}\label{eq:ky_gen}
    \left|k_xk_y-\cot(\theta)k_y^2\right|\approx \frac{z(\bar{r})}{h^2}
\end{align}
and
\begin{align}\label{eq:ks_gen}
    \left|k_x - \cot(\theta) k_y \right|\approx \frac{\bar{r}}{\ell_0}.
\end{align}

\begin{figure}
    \centering
    \subfigure[]{\includegraphics[width=0.45\textwidth]{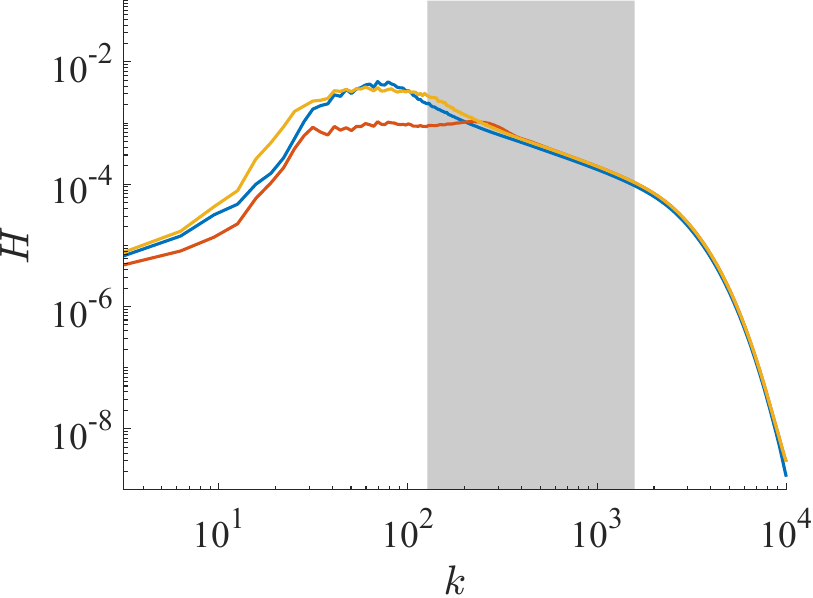}}
    \caption{The enstrophy spectrum corresponding to the vorticity field \eqref{eq:omega1} for different angles between the expanding and contracting directions: $\theta=90^\circ$ (blue), $\theta=45^\circ$ (yellow) and $\theta=11.3^\circ$ (red). The gray background represents the inertial range for $\theta=90^\circ$.}
    \label{fig:angle}
\end{figure}

The enstrophy spectrum in both the inertial and dissipation range changes rather insignificantly when the angle $\theta$ is varied. The most notable differences are at lower wavenumbers, as illustrated in \autoref{fig:angle}. In particular, the lower bound $k_c$ of the inertial range shifts towards higher wavenumbers when $\theta$ deviates increasingly from $90^\circ$. 
The shift of $k_c$ is due primarily to the spectral gap defined by Equation \eqref{eq:ks_gen} progressively overlapping with the hyperbolic region near the $k_x$ axis.

\bibliographystyle{jfm}
\bibliography{references}

\end{document}